\documentclass[useAMS,usenatbib,usegraphicx,aas_macros]{mn2e}
\usepackage{bm}
\usepackage{ifthen}
\usepackage{latexsym}

\newcommand{\calH}{\mathcal H}

\newcommand{\cm}{{\mathrm{cm}}}
\newcommand{\second}{{\mathrm{s}}}
\newcommand{\gram}{{\mathrm{g}}}
\def\Msun{\hbox{$M_\odot$}}

\newcommand{\BHphi}{\hat{\bm{\varphi}}}

\newcommand{\BHi}{\hat{\bm{\imath}}}
\newcommand{\BHj}{\hat{\bm{\jmath}}}
\newcommand{\BHk}{\hat{\bm{k}}}

\newcommand{\BHu}{\hat{\bm{u}}}
\newcommand{\BHw}{\hat{\bm{w}}}

\title[X-rays from two-temperature dipole accretion funnel]{
	X-ray emissions from two-temperature accretion flows
	within a dipole magnetic funnel
}
\author[Saxton et al.]{
Curtis J. Saxton$^{1}$,
Kinwah Wu$^{1,2}$,
Jo\~{a}o.~B.~G. Canalle$^{3}$,
Mark Cropper$^{1}$   
\newauthor
\& 
Gavin Ramsay$^{1,4}$   
\\
$^{1}${
Mullard Space Science Laboratory, University College London,
Holmbury St Mary, Dorking, Surrey RH5~6NT, England
}\\
$^{2}${
TIARA, Department of Physics, National Tsing Hua University,  
  Hsinchu 300, Taiwan  
}\\
$^{3}${
State University of Rio de Janeiro,
Rua S\~{a}o Francisco Xavier, 524/3023-D, CEP 20559-900,
Rio de Janeiro, RJ, Brazil
}\\ 
$^{4}${
Armagh Observatory, College Hill, Armagh BT61 9DG, N. Ireland
}\\ 
}
\begin{document}

\date{Accepted ----. Received ----; in original form ----}

\pagerange{\pageref{firstpage}--\pageref{lastpage}} \pubyear{2006}

\maketitle

\label{firstpage}

\begin{abstract}  
We investigate the hydrodynamics of
   accretion channelled by a dipolar magnetic field (funnel flows).
We consider situations in which the electrons and ions in the flow
  cannot maintain thermal equilibrium (two-temperature effects)
  due to strong radiative loss,
  and determine the effects on the keV X-ray properties of the systems.
We apply this model to investigate the accretion shocks of white dwarfs
   in magnetic cataclysmic variables.
We have found that the incorporation of two-temperature effects
   could harden the keV X-rays.
Also, the dipolar model yields harder X-ray spectra
   than the standard planar model
   if white dwarf is sufficiently massive ($\ga 1M_\odot$).
When fitting observed keV X-ray spectra of magnetic cataclysmic variables,
   the inclusion of two-temperature hydrodynamics
   and a dipolar accretion geometry
   lowers estimates for white-dwarf masses
   when compared with masses inferred from models excluding these effects.
We find mass reductions $\la 9\%$ in the most massive cases.
\end{abstract}


\section{Introduction}
\label{s.intro}

Field-channelled accretion occurs in a variety of stellar systems, 
   from young stellar objects 
   \citep[e.g.][]{koenigl1991,hartman1994,gullbring2000,lamzin2001,romanova2003,stelzer2004,gregory2006} 
   to compact stars 
   \citep[e.g.][]{elsner1977,ghosh1978,arons1993,lovelace1995,li1996b,li1999,kryukov2000,koldoba2002,toropina2003,canalle2005}, 
   where the magnetic stress of the accretor exceeds
   the ram pressure of the accretion flow.  
For accretion onto white dwarfs in magnetic cataclysmic variables (mCVs)
  \citep{warner1995,cropper1990}, 
   the magnetic field not only directs the flow 
   but could also dictate the radiative-loss processes
  \cite[see e.g.][]{wu2003}.   
This, in turn, alters the flow hydrodynamics,
   leading to observable consequences, 
   for example, in the optical/IR polarisation
   and keV X-ray spectra of the system.   
    
The keV X-ray emitting regions of mCVs are usually located 
   at the magnetic field footpoints on the white-dwarf surface.
Here an optically thin plasma slab heated by a strong shock, 
   at which the supersonic flows in the upstream region 
   decelerate abruptly to attain subsonic speeds.   
Typically, the shock temperature is  
   $T \approx 3 G M_{\rm w} m_{\rm H}/8 kR_{\rm w} \sim 10 - 50$~keV
   (where $G$ is the gravitational constant, 
   $k$ is the Boltzmann constant, 
   $m_{\rm H}$ is the hydrogen atomic mass, 
   $M_{\rm w}$ is the white-dwarf mass, 
   $R_{\rm w}$ is the white-dwarf radius). 
The shock-heated flow is cooled by emission 
    of free-free and line X-rays, and optical/IR cyclotron radiation.  

The post-shock regions of white dwarfs in mCVs
  are stratified in temperature and density, 
  and the flow structures are in general well described by Aizu-type models   
  \citep{aizu1973,chevalier1982,wu1994a,wu1994b,cropper1999,wu2000}.
These models assume that the electron and ions are in thermal equilibrium, 
   via collisions, 
   and share the same local temperature.  
(Hereafter, these models are denoted as standard one-temperature (1T) models.) 
The standard 1T models have successfully explained
   the line features in the keV X-ray spectra of mCVs.  
However using the standard 1T models to generate spectral fits for mCVs    
   tends to require white-dwarf masses systematically higher than 
   the masses measured by other techniques 
   \citep[see][]{ramsay1998,ramsay2000}. 
This has led to questions on the validity the 1T  approximation.            
    
It has long been recognised
   \citep{fabian1976,king1979,lamb1979}
   that the electrons and ions in the post-shock flows of mCVs 
   might be unable to maintain thermal equilibrium
   solely via electron-ion collisions
   \citep[see also][]{imamura.thesis,imamura1987,imamura1996,woelk1996,saxton1999,saxton2001,fischer2001}. 
The departure of the electron temperature
   from the ion temperature can be severe  
   if the radiative-loss timescale of the flow
   is much shorter than the dynamical timescale. 
In such situations,
   two-temperature (2T) effects on the X-ray spectra are no longer negligible.  
Two-temperature effects on X-ray spectra of mCVs were previously investigated
   assuming planar stratified flow 
   \citep{saxton2005}. 
However, the accretion flows in mCVs are not  planar. 
There is evidence that the accretion flows
   form curved funnel/curtain-like columns along the magnetic field lines 
   \citep{ferrario1996,heerlein1999}.  
The hydrodynamics in funnel flow and in planar are qualitatively different, 
  especially for tall post-shock columns
  where a variation of the gravitational potential is significant
  \citep{canalle2005}.  
          
Here we investigate 2T accretion in mCVs 
   where the flows are guided by the white-dwarf magnetic field geometry.   
This work is a generalisation of both the 2T planar flow calculations of 
  \cite{saxton2005}
  and the 1T dipole-field channelled flow calculations of 
  \cite{canalle2005}. 
We organise the paper as follows. 
In \S2 we present the model assumptions and the hydrodynamic formulation;  
  in \S3 we show the hydrodynamical structures and X-ray spectra 
     obtained from the calculations 
     and discuss the properties of the flows
     (and their dependence on system paramters).  
In \S4 we summarise the major findings.

\section{Hydrodynamic Formulation}

For a gas with adiabatic index $\gamma$,
density $\rho$,
velocity ${\mathbf v}$
and pressure $P$,
subject to a gravitational field ${\mathbf g}$,
the general hydrodynamic equations
for conservation of mass, momentum and energy
can be expressed as
\begin{equation}
	{{\partial\rho}\over{\partial t}}
	+\nabla\cdot\rho{\mathbf v}
	=0
	\ ,
\label{eq.vector.mass}
\end{equation}
\begin{equation}
	{\partial\over{\partial t}}\rho{\mathbf v}
	+\nabla\cdot\rho{\mathbf v}{\mathbf v}
	+\nabla P
	=\rho{\mathbf g}
	\ ,
\label{eq.vector.momentum}
\end{equation}
and
\begin{equation}
	\left({
		{\partial\over{\partial t}}+{\mathbf v}\cdot\nabla
	}\right) P
	-{{\gamma P}\over\rho}
	\left({
		{\partial\over{\partial t}}+{\mathbf v}\cdot\nabla
	}\right) \rho
	=-(\gamma-1)\Lambda
	\ ,
\label{eq.vector.energy}
\end{equation}
where $\Lambda$ is the volumetric energy loss rate
due to radiative cooling.
Emission by electrons far exceeds emission by the ions.
Unlike \cite{canalle2005},
who assumed temperature equilibrium between electrons and ions,
we will consider cases where electron radiative cooling
is significant compared to collisional heating by the ions.
This entails a separate energy equation for the electron sub-fluid,
\begin{equation}
	\left({
		{\partial\over{\partial t}}+{\mathbf v}\cdot\nabla
	}\right) P_{\rm e}
	-{{\gamma P_{\rm e}}\over\rho}
	\left({
		{\partial\over{\partial t}}+{\mathbf v}\cdot\nabla
	}\right) \rho
	=(\gamma-1)(\Gamma-\Lambda)
	\ ,
\label{eq.vector.electron}
\end{equation}
where $P_{\rm e}$ is the electron partial pressure.
The electron cooling function,
$\Lambda=\Lambda(\rho,P_{\rm e})$,
and the collisional heating function,
$\Gamma=\Gamma(\rho,P,P_{\rm e})$,
depend on local hydrodynamic variables
\citep{spitzer1962,rybicki,imamura1996}.

We consider accretion flows that are channelled by a dipolar magnetic field
centred on the accreting stellar object.
The coordinate $w$ measures paths
along a magnetic field line towards the accretor.
A transverse coordinate $u=\sin^2\theta_*$
is related to the colatitude of the accretion hot-spot ($\theta_*$).
The azimuthal angle $\varphi$ is the same as in cylindrical coordinates.
We seek stationary solutions for the flow structure,
so we omit temporal derivatives from 
(\ref{eq.vector.mass})--(\ref{eq.vector.electron}).
We recast the equations in curvilinear coordinates,
$(\BHu,\BHw,\BHphi)$.
For flows along the field line,
there is no velocity component in the $\BHu$ or $\BHphi$ directions.
Then the hydrodynamic equations for
mass flux, momentum flux,
total energy and electron energy
simplify to
\begin{equation}
	{\partial\over{\partial w}} h_1 h_3 \rho v =0
	\ \ \Rightarrow\ \ 
	h_1 h_3 \rho v = C
	\ ,
\end{equation}
\begin{equation}
	{v\over{h_2}}{{\partial v}\over{\partial w}}
	+{1\over{h_2\rho}}{{\partial P}\over{\partial w}}
	=g_w
	\ ,
\label{eq.curvy.momentum}
\end{equation}
\begin{equation}
	{v\over{h_2}}{{\partial P}\over{\partial w}}
	-{\gamma\over{h_2}}{{Pv}\over\rho}{{\partial\rho}\over{\partial w}}
	=-(\gamma-1)\Lambda
	\ ,
\end{equation}
\begin{equation}
	{v\over{h_2}}{{\partial P_{\rm e}}\over{\partial w}}
-{\gamma\over{h_2}}{{P_{\rm e}v}\over\rho}{{\partial\rho}\over{\partial w}}
	=-(\gamma-1)(\Lambda-\Gamma)
	\ ,
\end{equation}
where $C$ is a constant proportional to the mass flux onto the stellar surface,
$C=\dot{m} h_{1*}h_{3*}$.
The functions $h_1(u,w)$ and $h_3(u,w)$ are metric terms
   of the curvilinear coordinate system
   (see Appendix~\ref{app.coordinates}).
The term $g_w={\mathbf g}\cdot\BHw$
   is gravitational acceleration along the magnetic field line
   (i.e. in the $w$ direction).

As expressed in
\cite{saxton2005},
the rate of energy exchange from ions to electrons
has the form
\begin{equation}
	\Gamma=X\rho^{5/2} P_{\rm e}^{-3/2} (P-\chi P_{\rm e})
	\ ,
\end{equation}
where $X$ and $\chi$ are numerical constants
that depend on the plasma composition.
The volumetric cooling rate of gas due to thermal bremsstrahlung radiation
is approximated as
\begin{equation}
	\Lambda_{\rm br}=A\rho^2\sqrt{{P_{\rm e}}\over\rho}
	\ ,
\end{equation}
with $A$ being another constant depending on composition.
\cite{saxton2005}
gave derivations of $A$, $X$ and $\chi$
and their values for a plasma of solar metallicity.
We assume a total cooling function
that includes bremsstrahlung and cyclotron contributions,
\begin{equation}
	\Lambda=\Lambda_{\rm br}+\Lambda_{\rm cy}
	=\Lambda_{\rm br}\left[{
		1 + \epsilon_{\rm s} f_{\rm cy}(\rho,P_{\rm e})
	}\right]
	\ ,
\end{equation}
with a cyclotron/bremsstrahlung emissivity ratio
given by the function $f_{\rm cy}$
defined by \cite{wu1994a,wu1994b,saxton1997},
refined by \cite{saxton1999,cropper1999,saxton2005}
and extended to the dipolar accretion model
in \cite{canalle2005}.

We rewrite the flow density, total pressure and electron partial pressure
in terms of the $h$ functions,
the flow constant $C$
and variables that have the dimensions of velocity,
\begin{equation}
	\rho={{C}\over{h_1h_3 v}}
	\ ,
\end{equation}
\begin{equation}
	P={{C}\over{h_1h_3}}(\xi-v)
	\ ,
\end{equation}
\begin{equation}
	P_{\rm e}={{C}\over{h_1h_3}}(\xi-v)
	\left({
		{\sigma\over{\sigma+1}}
	}\right)
	\ ,
\label{eq.def.Pe}
\end{equation}
where $\xi$ measures the ratio of momentum to mass fluxes
\citep[introduced in][]{cropper1999}
and $\sigma\equiv P_{\rm e}/P_{\rm i}$,
the electron to ion partial pressure ratio.
For the sake of further brevity,
we define
\begin{equation}
	p\equiv  {{h_1h_3}\over{C}} P_{\rm e}
	=
	{\sigma\over{\sigma+1}} (\xi-v)
	\ .
\end{equation}
We separate the finite and $v^{-1}$ terms
in the electron-ion energy exchange function
and the bremsstrahlung radiative cooling functions,
\begin{equation}
	\Gamma=
	X\left({
	{{C}\over{h_1h_3}}
	}\right)^2
	{{ \sigma + 1 -\chi\sigma }\over{ \sigma p^{1/2} v^{5/2} }}
	\equiv
	{C\over{h_1h_2h_3}}\hat\Gamma v^{-5/2}\ ,
\end{equation}
\begin{equation}
	\Lambda_{\rm br}=
	A\left({
		{{C}\over{h_1h_3}}
	}\right)^2
	p^{1/2} v^{-3/2}
	\equiv
	{C\over{h_1h_2h_3}}\tilde\Lambda v^{-3/2}
	\ .
\end{equation}
The functions
$\hat\Gamma\equiv h_1h_2h_3\Gamma v^{5/2}/C$
and
$\tilde\Lambda\equiv h_1h_2h_3\Lambda v^{3/2}/C$
lack explicit $v$ dependencies
and are finite everywhere.

As in
\cite{canalle2005},
the equations for momentum and total energy provide
\begin{equation}
	{{d\xi}\over{dw}}
	=
	{{g_wh_2}\over{v}}
	+\calH(\xi-v)
	\ ,
\label{eq.dxidw}
\end{equation}
\begin{equation}
	{{dv}\over{dw}}
	=
	{{
		-\left[{
			(\gamma-1)\tilde\Lambda v^{-3/2}
			+\gamma\calH (\xi-v)v
			+h_2 g_w
		}\right]
	}\over{
		\gamma(\xi-v)-v
	}}
	\ .
\label{eq.dvdw}
\end{equation}

Definition (\ref{eq.def.Pe}),
equations (\ref{eq.curvy.momentum})
and (\ref{eq.dvdw}) yield
\begin{equation}
	{{dP_{\rm e}}\over{dw}}
	=
	{{P_{\rm e} }\over{\xi-v} }
	\left({
		{{d\xi}\over{dw}}-{{dv}\over{dw}}
	}\right)
	-\calH P_{\rm e}
	+{{(P-P_{\rm e})^2 }\over{P} }
	{{d\sigma}\over{dw}}
	\ ,
\label{eq.dPedw}
\end{equation}
\begin{equation}
	{{d\sigma}\over{dw}}
	=-{{
		(\gamma-1)(\sigma+1)
	}\over{
		(\xi-v)v^{7/2}
	}}
	\left[{
		v\tilde\Lambda-(\sigma+1)\hat\Gamma
	}\right]
	\ .
\label{eq.dsigmadw}
\end{equation}

Strong shock conditions apply at the outer boundary,
while the condition $v=0$ applies at the stellar surface ($r=1$).
Our particular method of numerical solution is described in
Appendix~\ref{app.solver}.
We adopt the white dwarf mass-radius relation of \cite{nauenberg1972}.

\section{Results}

\subsection{Structure of the post-shock flow}

Table~\ref{table.basics}
summarises the basic properties of several models of mCVs,
with nearly direct (approximately radial) accretion onto a magnetic pole
(colatitude $\theta_*=0.001^\circ$)
and hot-spot area $10^{15}~\cm^2$.
We consider white dwarfs in the mass range $0.7 - 1.2 M_\odot$,
with surface magnetic field strengths $B_*\leq 50$~MG.
The ratio of electron to ion temperatures at the shock,
$(T_{\rm e}/T_{\rm i})_{\rm s}$,
is treated as a free parameter
\citep[e.g. as in][]{imamura1996,saxton1999,saxton2005}.
This ratio has little effect on the shock height,
$R_{\rm s}$.
This implies that the shock position is insensitive to
the details of heating at the shock precursor,
but depends mainly on
the white-dwarf's properties,
the inflow rate and field geometry.

\begin{figure*}
\begin{center}
\begin{tabular}{cc}
\includegraphics[width=8.5cm]{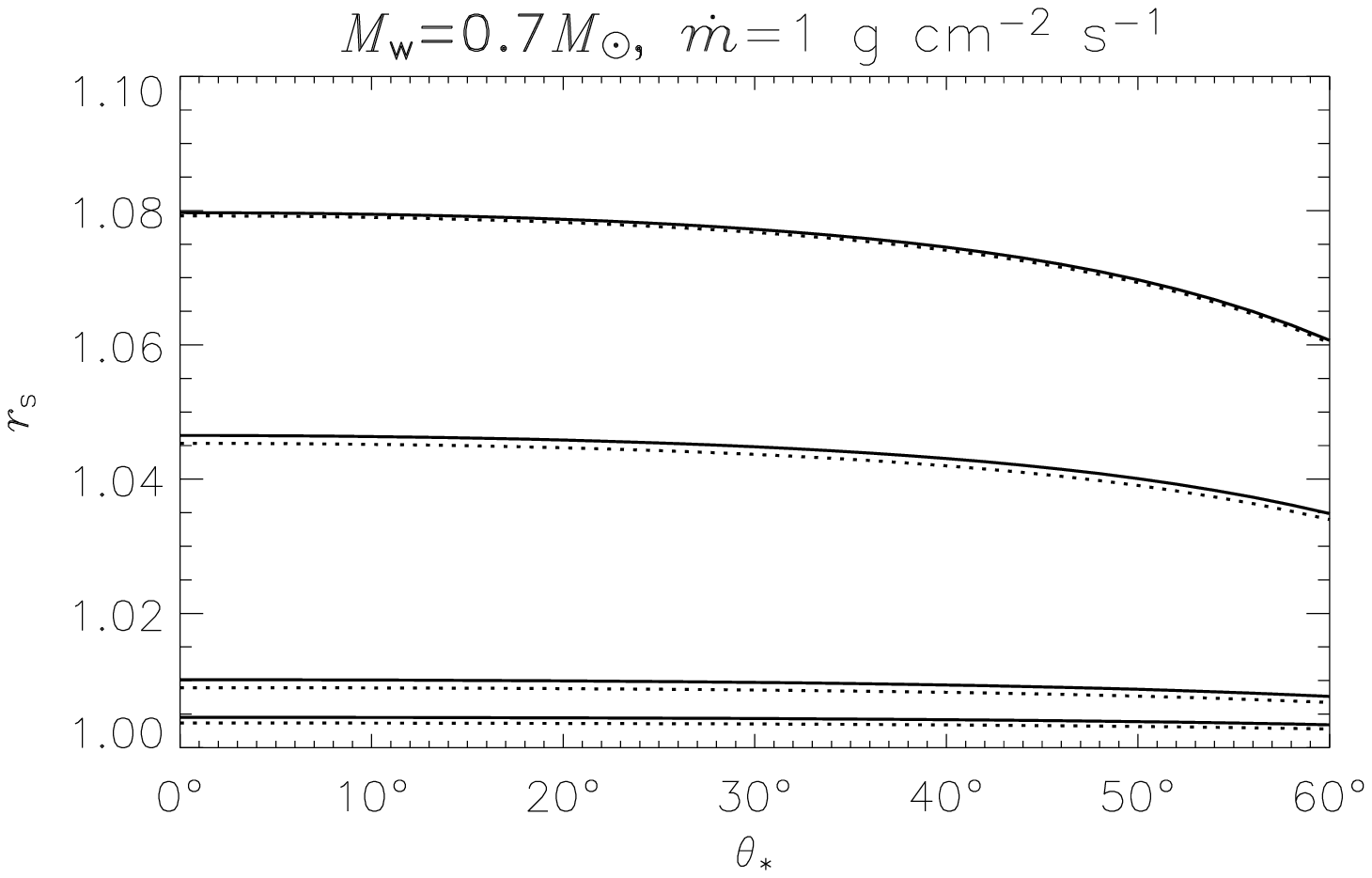}
&
\includegraphics[width=8.5cm]{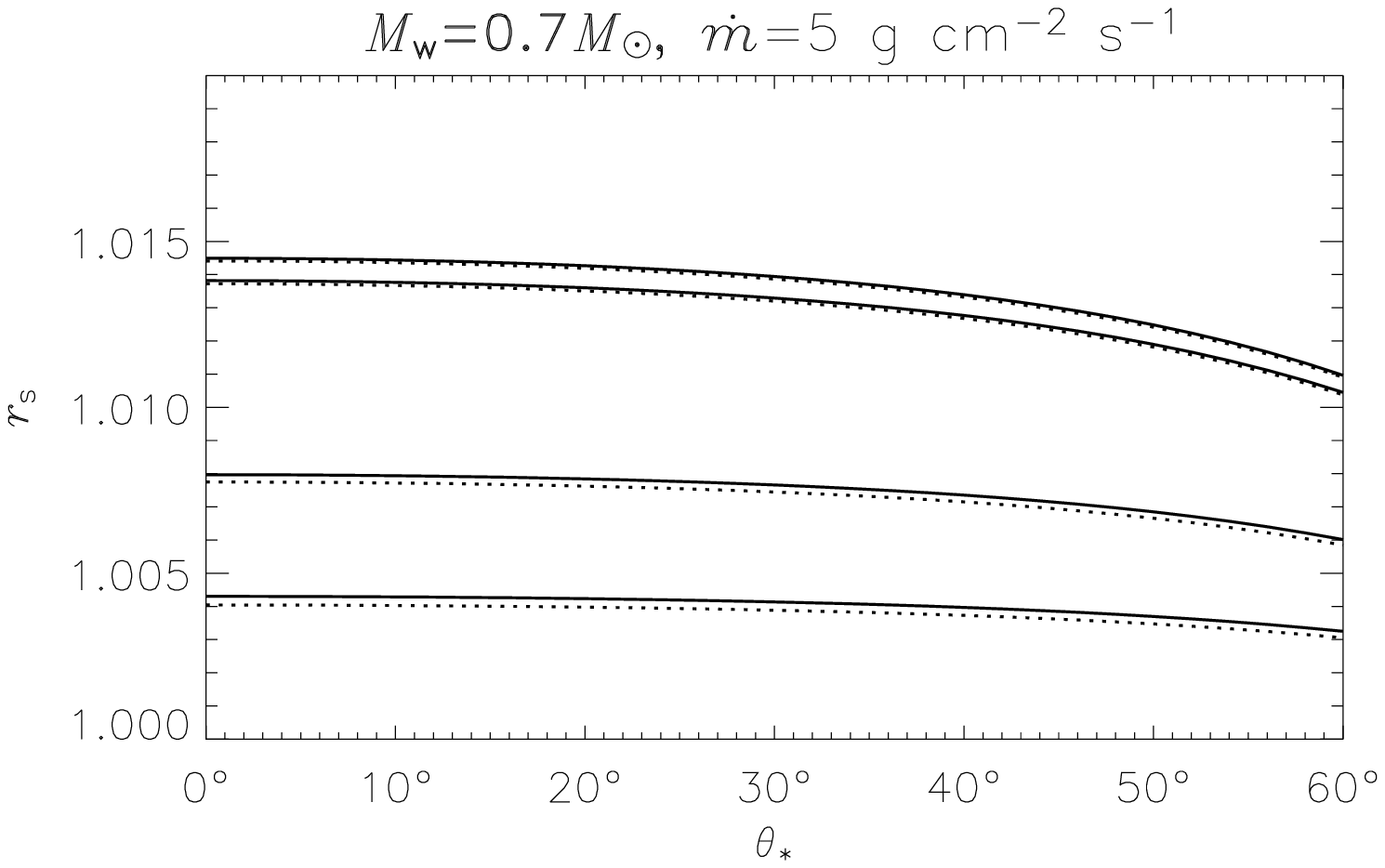}
\\
\includegraphics[width=8.5cm]{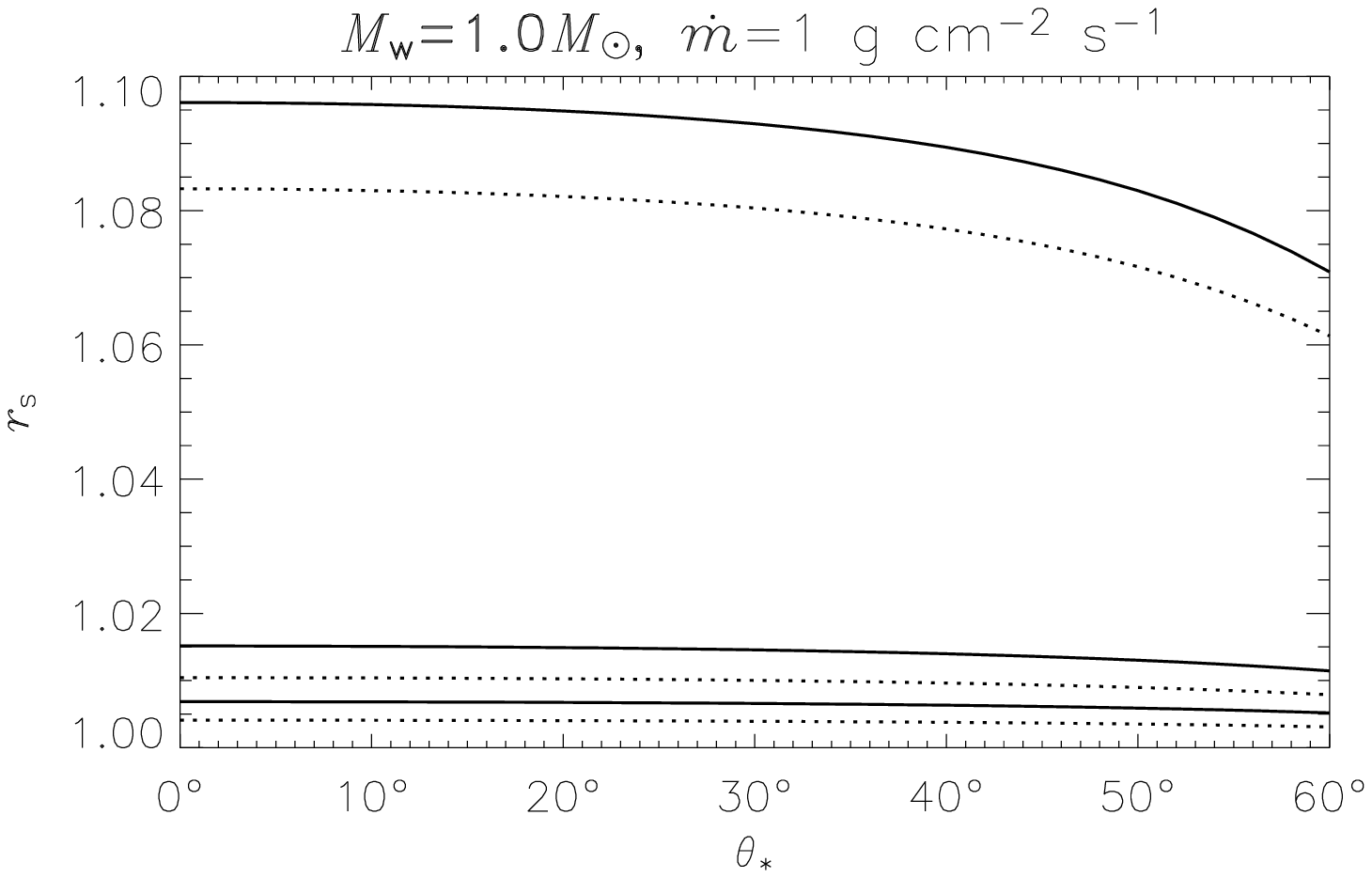}
&
\includegraphics[width=8.5cm]{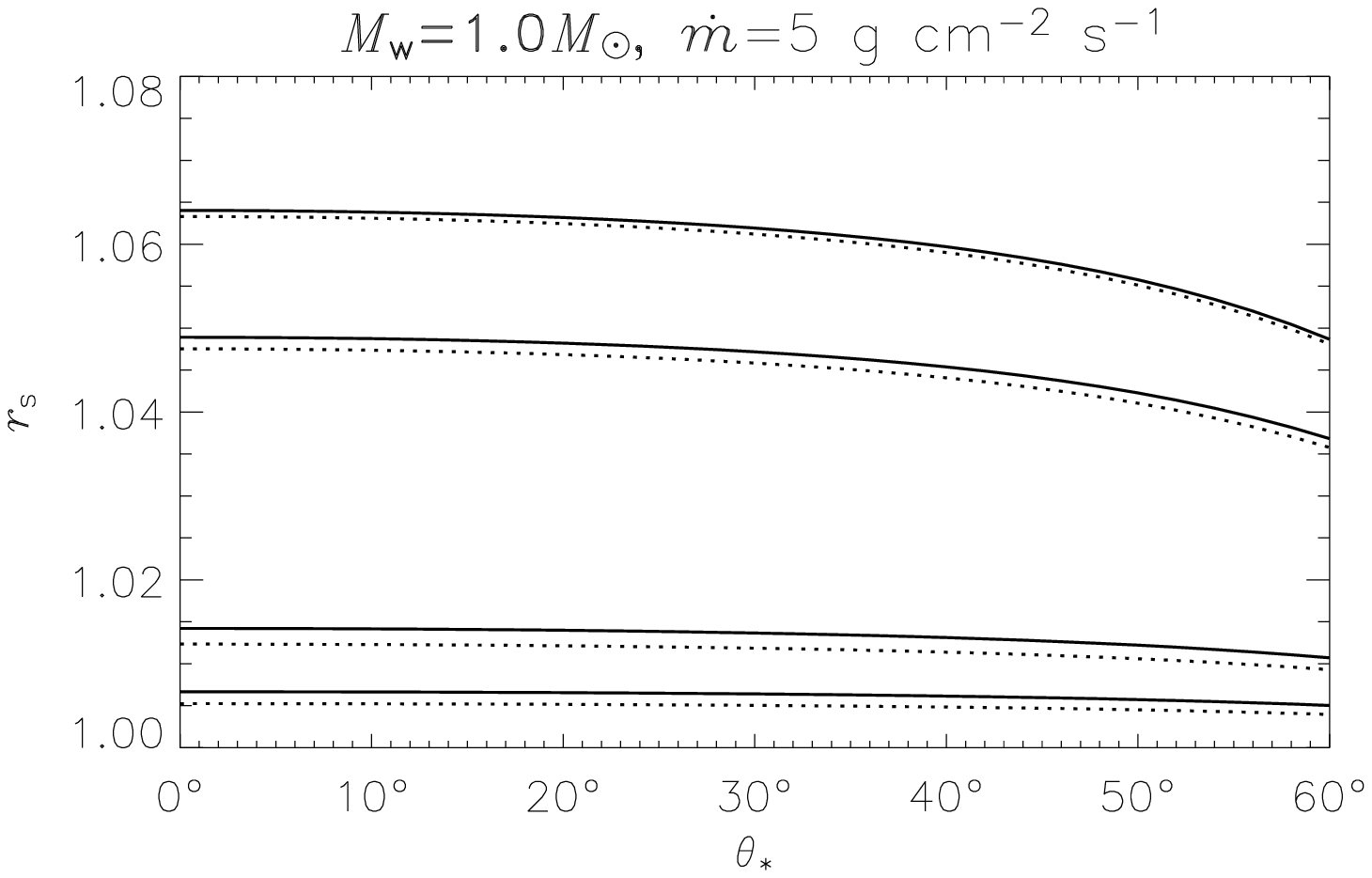}
\\
\includegraphics[width=8.5cm]{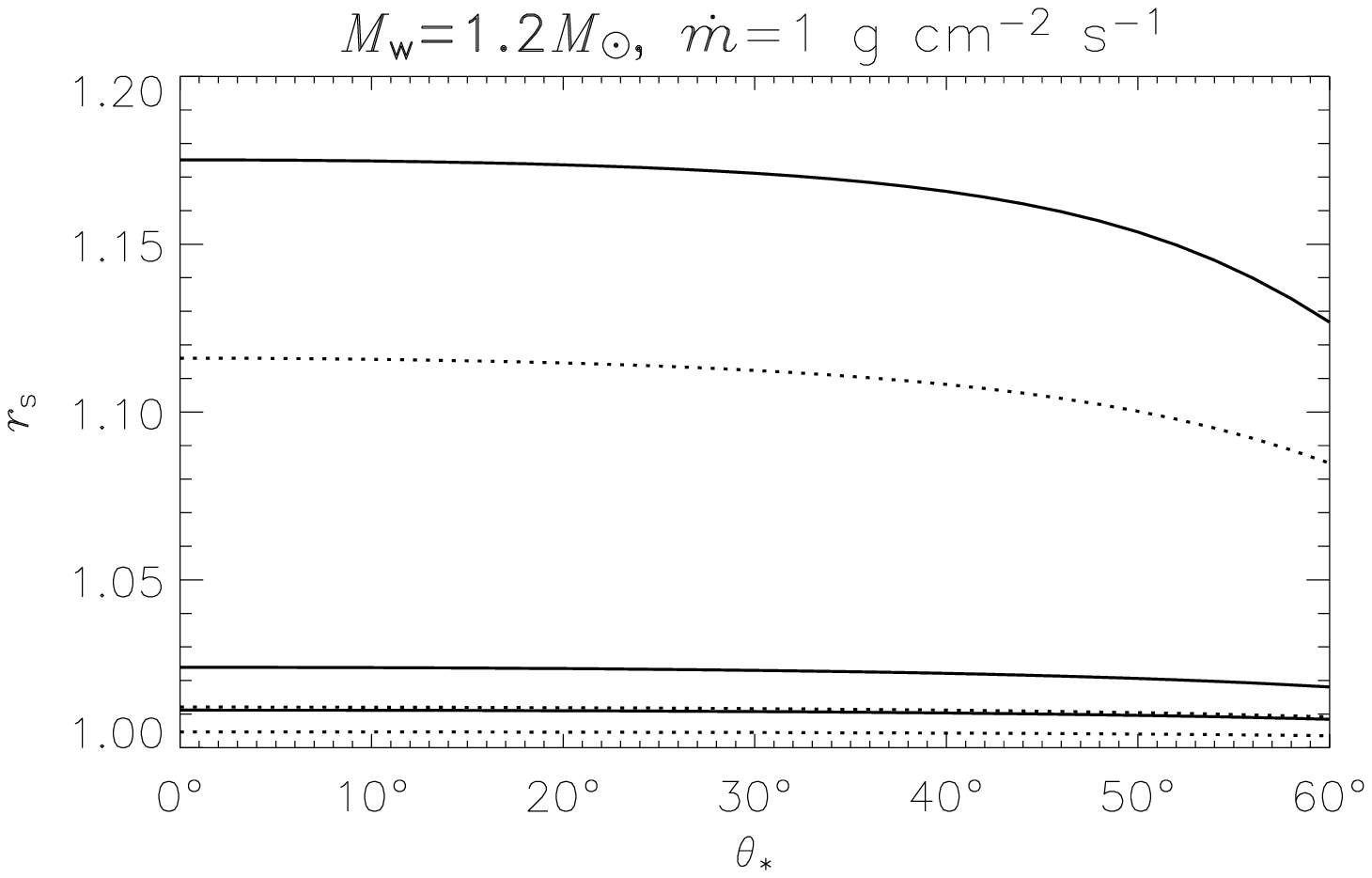}
&
\includegraphics[width=8.5cm]{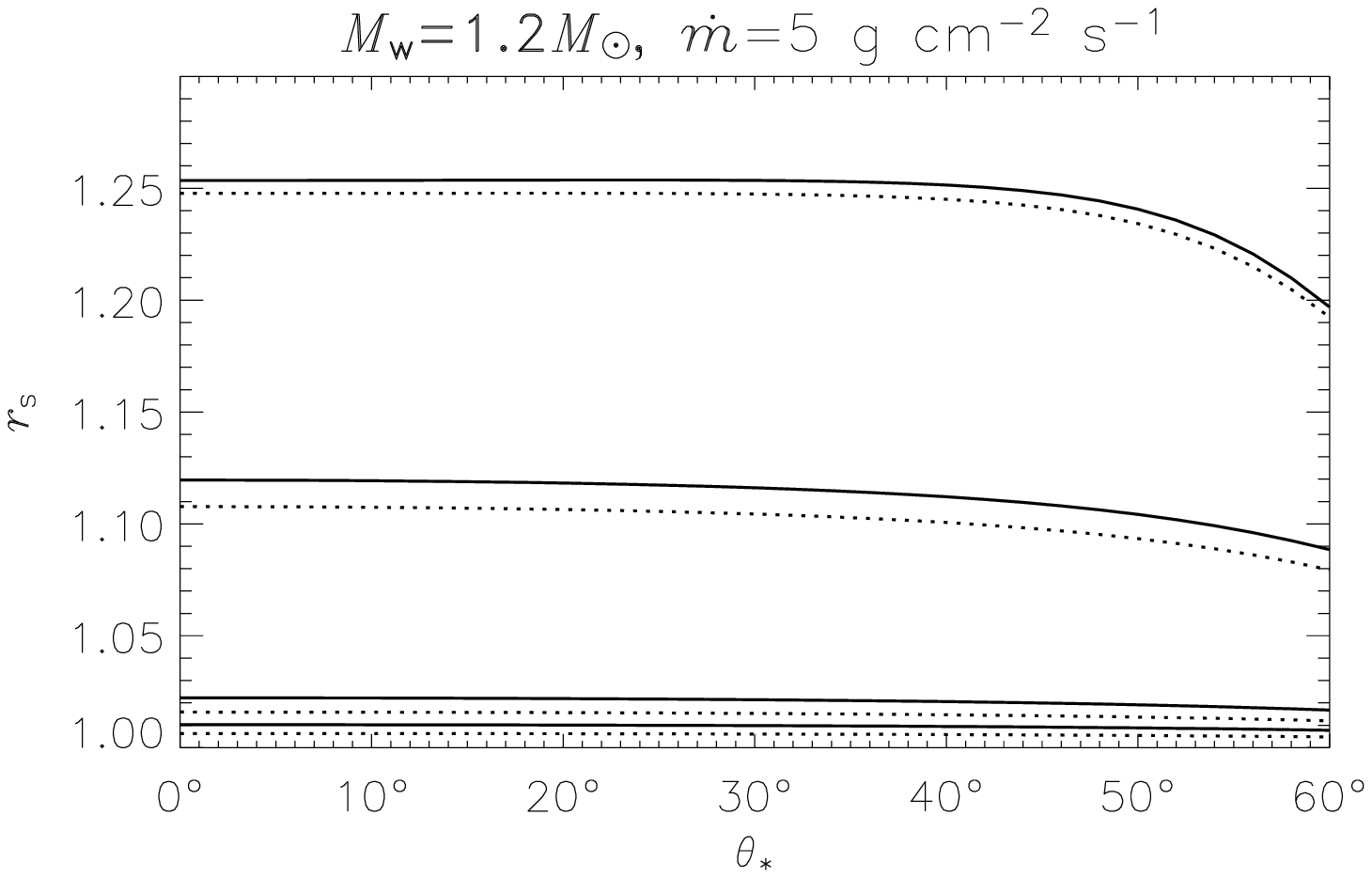}
\end{tabular}
\end{center}
\caption{
Effect of the accretion colatitude ($\theta_*$)
upon the shock radius,
$r_{\rm s}\equiv R_{\rm s}/R_{\rm w}$ in units of the stellar radius,
for cases with different specific accretion rates
($\dot{m}$),
white-dwarf mass ($M_{\rm w}$)
and magnetic field ($B_*$).
The accretion hot-spot area is $10^{15}~{\rm cm}^2$.
Solid lines mark the results of the 2T model;
dotted lines denote the corresponding results of the 1T model.
Within each panel,
paired curves from bottom to top
represent cases with field strengths
$B_* = 50, 30, 10$~MG;
a fourth pair shows cases with neglible cyclotron cooling
(yet channelled by a dipole funnel).
These low-field cases are insoluble when
$\dot{m}=1~\gram~\cm^{-2}~{\rm s}^{-1}$.
and $M_{\rm w}=1.0, 1.2\Msun$.
}
\label{fig.rshock}
\end{figure*}

Figure~\ref{fig.rshock}
shows how the shock radius varies
with the accretion colatitude $\theta_*$,
the magnetic field strength
and accretion rate.
These illustrative models assume that $T_{\rm e}=T_{\rm i}$ at the shock,
and accretion hot-spot area of $10^{15}~\cm^2$.
In all calculations
\citep[as in][]{canalle2005}
the shock is closer to the stellar surface
when $\dot{m}$ is greater.
The greatest variation of $R_{\rm s}$ with $\theta_*$
occurs in systems where the shock radius is large compared to $R_{\rm w}$.
We find that the 1T model
   (dotted curves) predicts lower shock positions
   than the 2T model does.
The difference is proportionally greater in large-$B_*$ cases,
   where the shock is typically nearer the stellar surface.
As the 2T model predicts
   increased $R_{\rm s}$ values,
   the shock temperatures are lower,
   $T_{\rm s}\propto R_{\rm s}^{-1}$.
This effect is greatest for large $M_{\rm w}$.

\begin{figure*}
\begin{center}
\begin{tabular}{c}
\includegraphics[width=17cm]{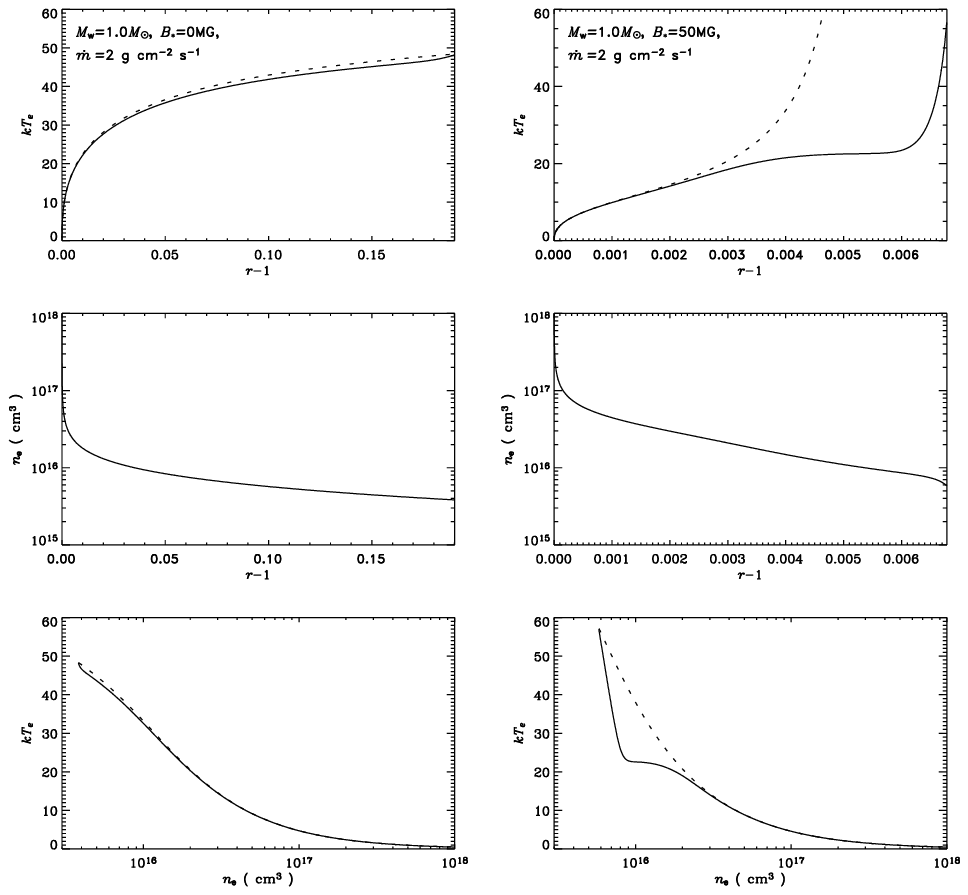}
\end{tabular}
\end{center}
\caption{
Profiles of electron temperature (upper row)
and
electron number density (middle row)
for accretion onto $1.0~M_\odot$ white dwarfs,
as a function of height above the surface.
The left and right columns respectively show cases
with negligible cyclotron cooling,
and with dominant cyclotron cooling.
The accretion spot is at the pole 
($\theta_*=0^\circ$).
Solid lines mark the results of the 2T model;
dotted lines denote the results of the 1T model.
Electron and ion temperatures match at the shock and stellar surface,
but the electrons are cooler elsewhere.
The bottom row shows the temperature/density phase structure:
in high-field cases the 2T model
has lower densities in the hotter strata.
}
\label{fig.structure2}
\label{fig.phase}
\end{figure*}

Figure~\ref{fig.structure2}
   portrays the radial distribution of electron density
   ($n_{\rm e}$) and temperature ($kT_{\rm e}$)
   in accretion models with $\theta_*=0^\circ$,
   $M_{\rm w}=1.0 M_\odot$
   and
   $\dot{m}=2~{\rm g}~{\rm cm}^{-2}~{\rm s}^{-1}$.
The 1T and 2T models predict 
   similar density and thermal structures in the lower field cases
   ($B_*=0$~MG).
However for stronger magnetic fields
   ($B_*=50$~MG)
   the models predict considerably different shock heights
   and qualitatively different density and temperature distributions.
In 2T structures,
   the hottest electrons 
   are confined to a narrower region near the shock
   than in 1T shocks with equivalent system parameters.
Plotting the $T_{\rm e}$ versus $n_{\rm e}$
   distributions,
   as in the lower panels of Figure~\ref{fig.phase},
   shows that in 2T models with strong magnetic fields
   the density of the hottest gas is lower than in 1T models.
This will have consequences for the post-shock X-ray emissions.

\begin{table}
\centering
\caption{
Parameters and properties of a set of representative accretion shocks,
varying the electron to ion temperature ratio at the shock
$(T_{\rm e,s}/T_{\rm i,s}$).
The white-dwarf surface magnetic field is given by $B_7=B_*/10$~MG,
and the specific accretion rate $\dot{m}$ is in units of g~cm$^{-2}$~s$^{-1}$.
If the magnetic field at the hot-spot is $B_*$
then its magnitude at the pole is
$B_{\rm p}=2B_*/\sqrt{4-3u}$.
}
\label{table.basics}
\begin{center}
\begin{tabular}{rrrrrrrrrrrrrrr}
\hline
\multicolumn{2}{c}{$T_{\rm e,s}/T_{\rm i,s}$~:}
&&&&1.0&0.5\\
$M_{\rm w}$&$B_7$&$\dot{m}$&$\theta_*$&$a_{15}$
&$r_{\rm s}$
&$r_{\rm s}$
\\
\\
\hline
0.7	&0.0	&1.0	&0.0	&1.0	&1.081	&1.080
\\
0.7	&1.0	&1.0	&0.0	&1.0	&1.047	&1.047
\\
0.7	&3.0	&1.0	&0.0	&1.0	&1.010	&1.011
\\
0.7	&5.0	&1.0	&0.0	&1.0	&1.005	&1.005
\\
\\
1.0	&0.0	&1.0	&0.0	&1.0	&1.610	&1.617
\\
1.0	&1.0	&1.0	&0.0	&1.0	&1.096	&1.103
\\
1.0	&3.0	&1.0	&0.0	&1.0	&1.015	&1.017
\\
1.0	&5.0	&1.0	&0.0	&1.0	&1.007	&1.008
\\
\\
1.2	&1.0	&1.0	&0.0	&1.0	&1.175	&1.206
\\
1.2	&3.0	&1.0	&0.0	&1.0	&1.024	&1.028
\\
1.2	&5.0	&1.0	&0.0	&1.0	&1.011	&1.014
\\
\\
\hline
\end{tabular}
\end{center}
\end{table}

\subsection{X-ray emissions}

For each of the solutions for the post-shock accretion flow,
   we calculate an X-ray spectrum
   following the methods of
   \cite{cropper1999} and \cite{saxton2005}.
The post-shock volume is divided into $\ga 10^4$ strata
   according to the spatial steps of the numerically integrated flow profile.
The volume of each stratum is inferred from the step size $dw$,
   the hot-spot area
   and the crossectional area scaling factor $h_1h_3/h_{1*}h_{3*}$.
The local electron density and temperature 
   $(n_{\rm e},T_{\rm e})$
   are obtained from the dimensionless flow model
   and the scaling units, $(R_{\rm w},V_{\rm w},C)$.
We assume that the X-ray line and continuum emission are optically thin,
   and invoke the {\sf XSPEC} implementation of the MEKAL thermal plasma model
   \citep{mewe1985,kaastra1993}
   to calculate emissivities for local
   $(n_{\rm e},T_{\rm e})$
   values.
Volumetric integration over all the post-shock strata
   yields a synthetic X-ray spectrum.
We exclude the densest strata,
   with $n_{\rm e}\geq 10^{18}~\cm^{-3}$,
   which the spectral model cannot handle
   (and which in any case represent optically thick layers
   merging into the atmosphere of the white dwarf).
We omit the effect of shadowing by a white dwarf
   which obscures its own accretion inflow.

\cite{saxton2005}
studied 2T effects in a planar accretion model,
without the effects of gravity
and varying width of the magnetic accretion funnel.
They found that 2T effects change the keV X-ray spectra
most significantly for cases with larger $M_{\rm w}$.
For given system parameters,
the 2T model gave a harder continuum
but relatively weaker line emission.
The ratio of electron to ion temperatures at the shock
was found to have little effect on the spectra.
This was explained in terms of the concentration of X-ray emission
near the stellar surface
(well downstream from the shock)
by which point collisional energy exchanges
have nearly equilibrated the electrons and ions.

\begin{figure*}
\begin{center}
\begin{tabular}{ccc}
\includegraphics[width=5.5cm]{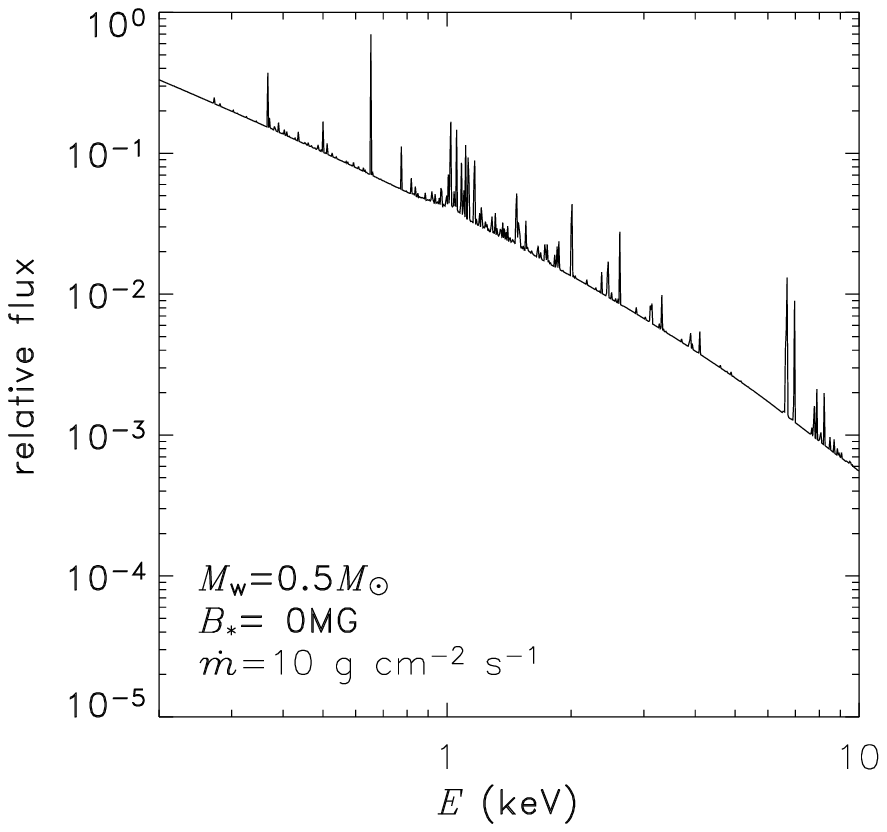}
&
\includegraphics[width=5.5cm]{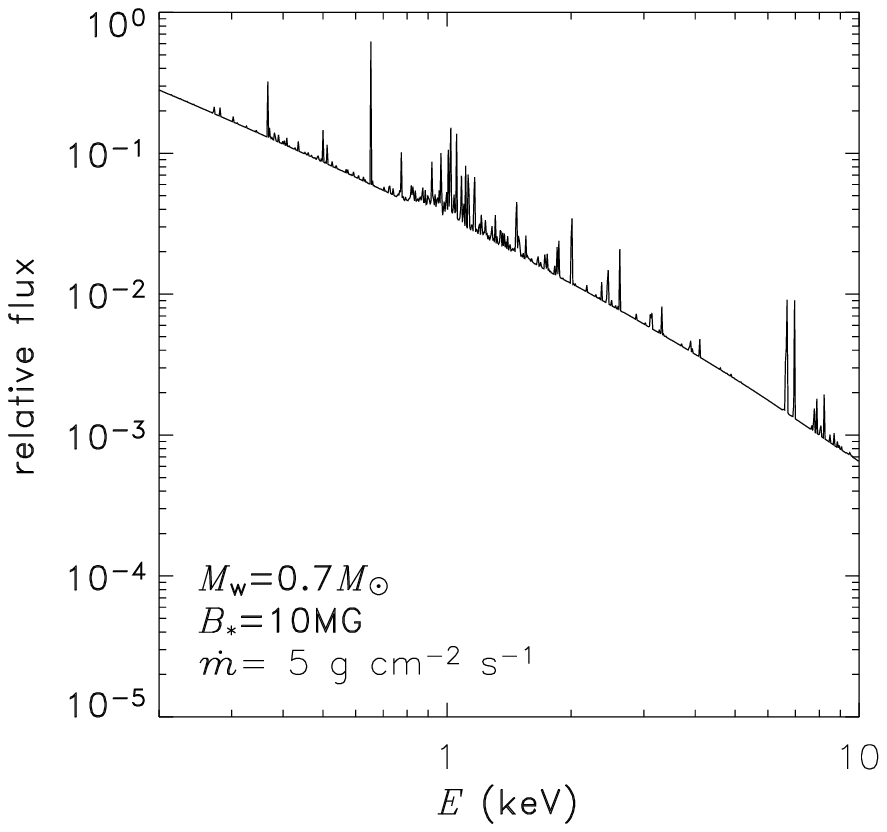}
&
\includegraphics[width=5.5cm]{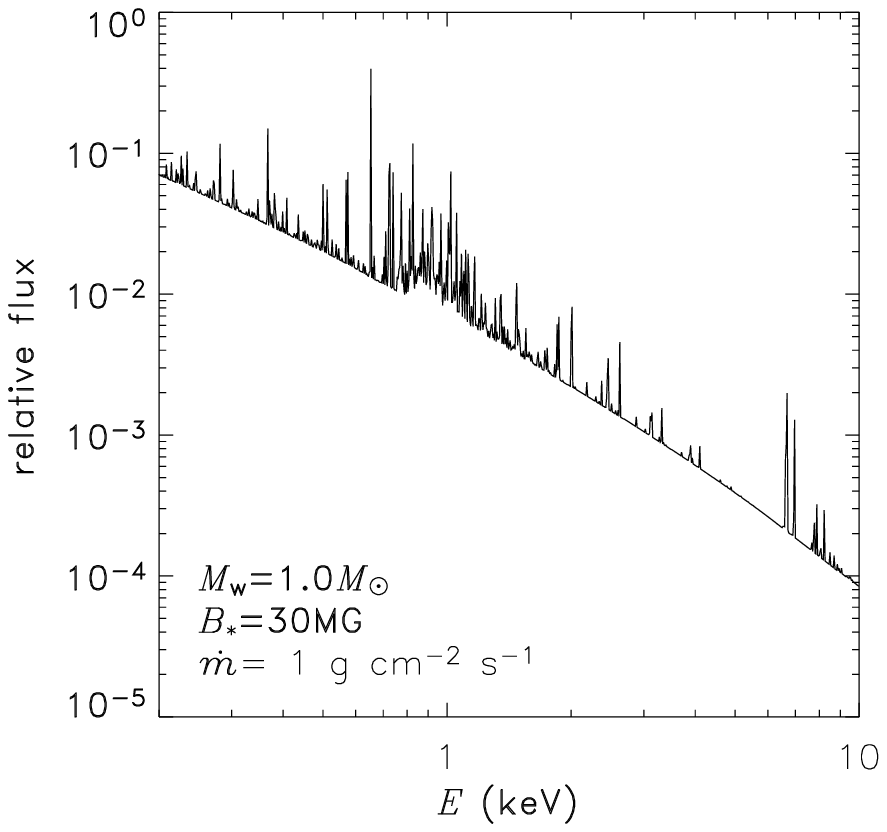}
\end{tabular}
\end{center}
\caption{ 
X-ray spectra for photon energies $0.2 - 10$~keV,
calculated for dipole-field channelled 2T accretion
onto white dwarfs with parameters as indicated in the panels.
} 
\label{fig.xray.typical}
\end{figure*}

Our present calculations,
   with a dipolar accretion funnel and 2T effects,
   typically produce spectra such as those in
Figure~\ref{fig.xray.typical}.
The slope of the continuum is steep,
but there are variations in the curvature of the continuum
and the details of line strengths.
The effects of the various system parameters
are best illustrated by plots of the ratios
of spectra calculated with different conditions or assumptions.

\subsubsection{Dipolar vs planar accretion}

\begin{figure*}
\begin{center}
\begin{tabular}{c}
\includegraphics[width=17.7cm]{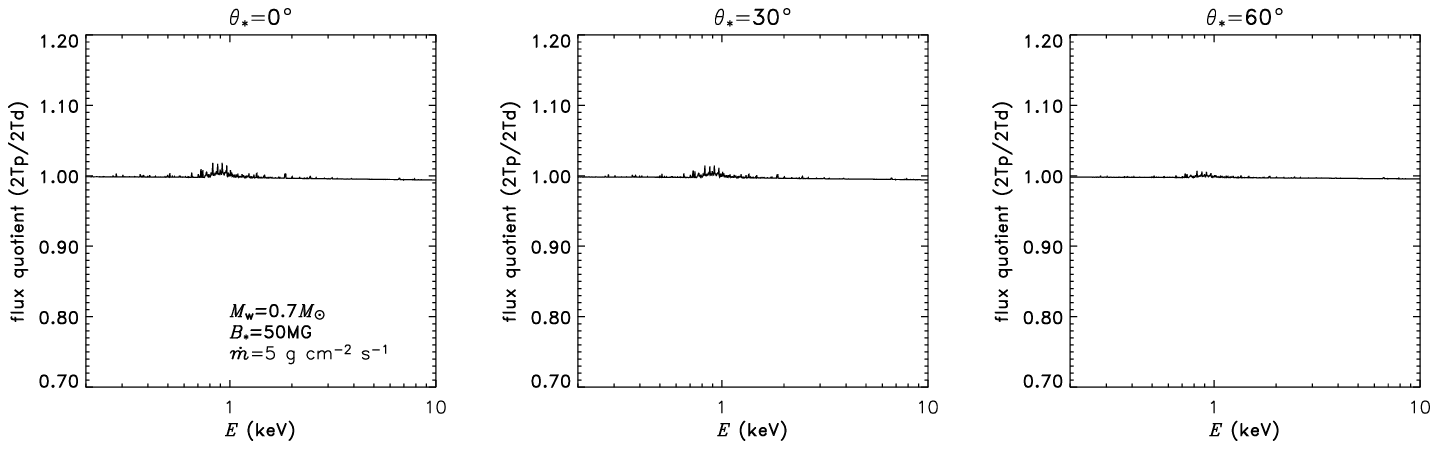}
\\
\includegraphics[width=17.7cm]{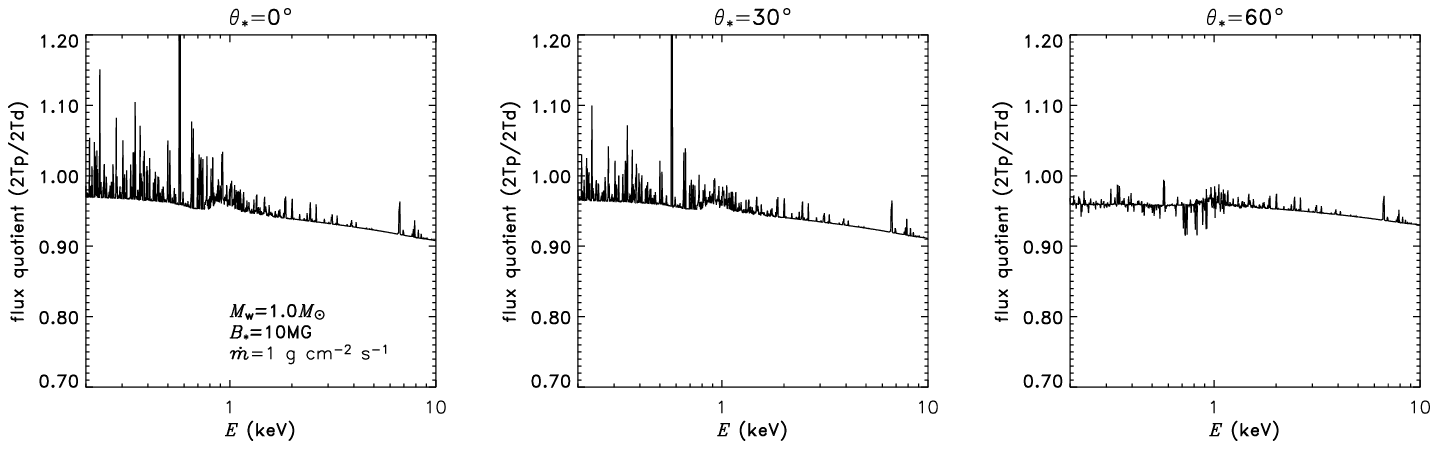}
\end{tabular}
\end{center}
\caption{
Quotients of spectra in 2T planar accretion model 
  to spectra in the corresponding dipolar accretion model
  (2Tp/2Td).
The columns show different cases of the accretion colatitude,
   $\theta_*=0^\circ, 30^\circ, 60^\circ$.
The top row represents a system in which the shock height is small 
   compared to the stellar radius.
The bottom row shows a system with a relatively high shock.  
In the latter,
   spectra differ significantly between the dipolar and planar models.
}
\label{fig.compare.flat}
\end{figure*}

Figure~\ref{fig.compare.flat}
shows the ratios of X-ray spectra calculated in
2T models with identical system parameters
but dipolar verses planar accretion geometry.
In all the cases we calculated,
the dipolar model provides a harder continuum,
and reduced line emission,
particularly for photon energies $\la 1$~keV.
These effects are small for a white dwarf of $0.7\Msun$,
but are more considerable for $1.0\Msun$ and more massive cases.
The dipolar (2Td) and planar (2Tp) models are more in agreement
regarding continuum at photon energies $\la 1$~keV
if the accretion hot-spot is at high $\theta_*$
(further from the magnetic pole).

\subsubsection{Two-temperature vs one-temperature dipolar inflows}

\begin{figure*}  
\begin{center}  
\begin{tabular}{c}  
\includegraphics[width=17.7cm]{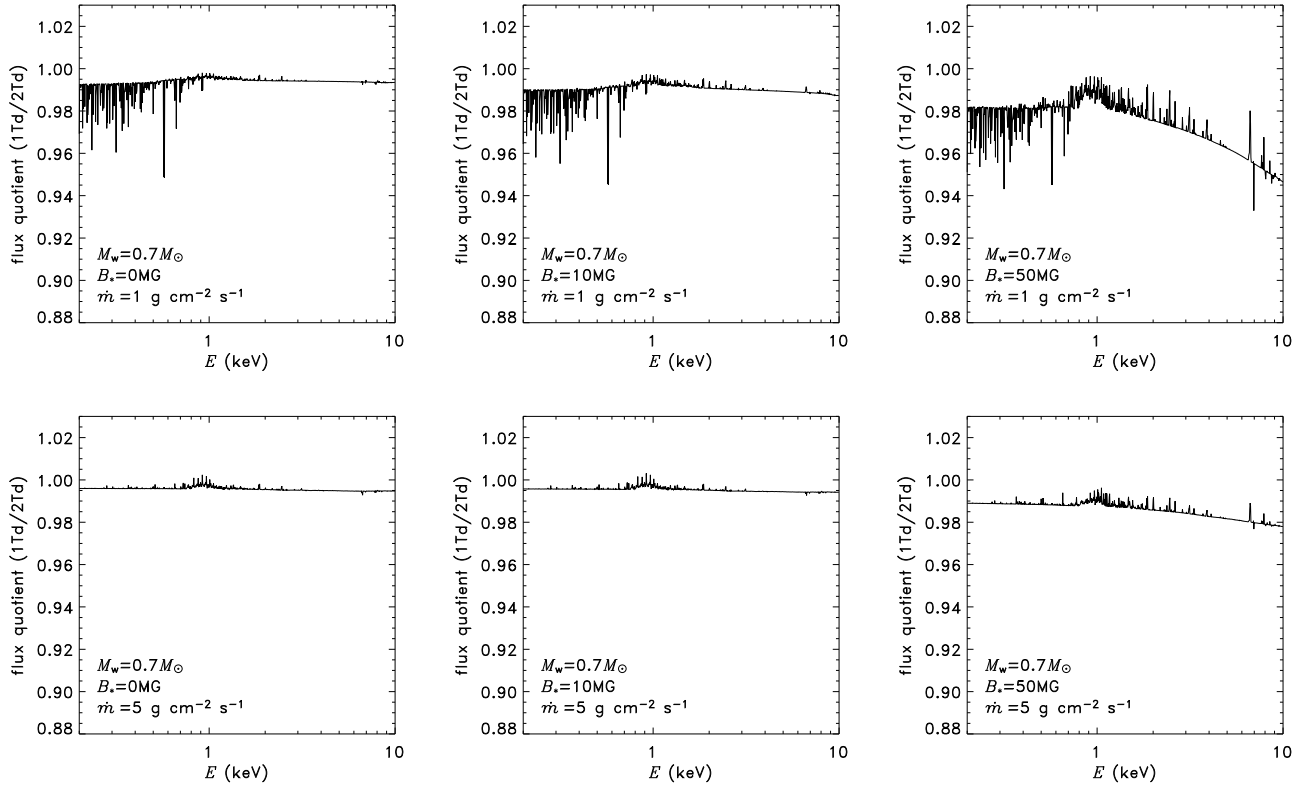} 
\end{tabular}  
\end{center}  
\caption{    
Quotients of spectra in 1T and 2T dipolar models (1Td/2Td).
In these cases the  white-dwarf mass is $0.7~M_\odot$
and accretion occurs onto the pole ($\theta_*=0^\circ$).
}  
\label{fig.xquot.07}
\end{figure*}

Figure~\ref{fig.xquot.07} shows the quotients of X-ray spectra
over the $0.2-10$~keV band,
from 1T and 2T dipolar calculations
for a white-dwarf mass of $0.7\Msun$.
The X-ray continuum is generally brighter in a 2T model.
In cases where bremsstrahlung is the dominant cooling process
($B_*=0$~MG)
the X-ray continuum has equivalent spectral slopes
in 1Td and 2Td models,
up to photon energies $\sim5$~keV.
Above this energy, the 1Td model
predicts a marginally harder continuum.
The 1Td models predict $\sim 0.5\%$  stronger emission lines
around 1~keV, but lines around $\sim 7$~keV are weakened by a similar degree.

For non-zero but small magnetic fields
($B_*=10$~MG)
the shape of the continuum is similar to the zero-field cases,
but the 1Td and 2Td calculations
yield greater discrepancies about the line strengths,
up to $\sim 1\%$.
The specific accretion rate, $\dot{m}$,
has a slightly greater effect on the disagreements between the predicted lines.

For a stronger magnetic field
($B_*=30$~MG)
the 1Td model yields a softer continuum than
equivalent 2Td calculations.
The 1Td model generally predicts stronger emission lines,
by up to $\sim 2\%$ for the cases with lowest $\dot{m}$.
Two-temperature effects are less apparent in the keV X-ray continuum and lines
when the accretion rate is larger.
The spectral effects generally strengthen as the magnetic field increases;
the greatest extreme in our plots is the case with
$B_*=50$~MG and $\dot{m}=1~\gram~\cm^{-2}~\second^{-1}$.

\begin{figure*}
\begin{center}
\begin{tabular}{c}
\includegraphics[width=17.7cm]{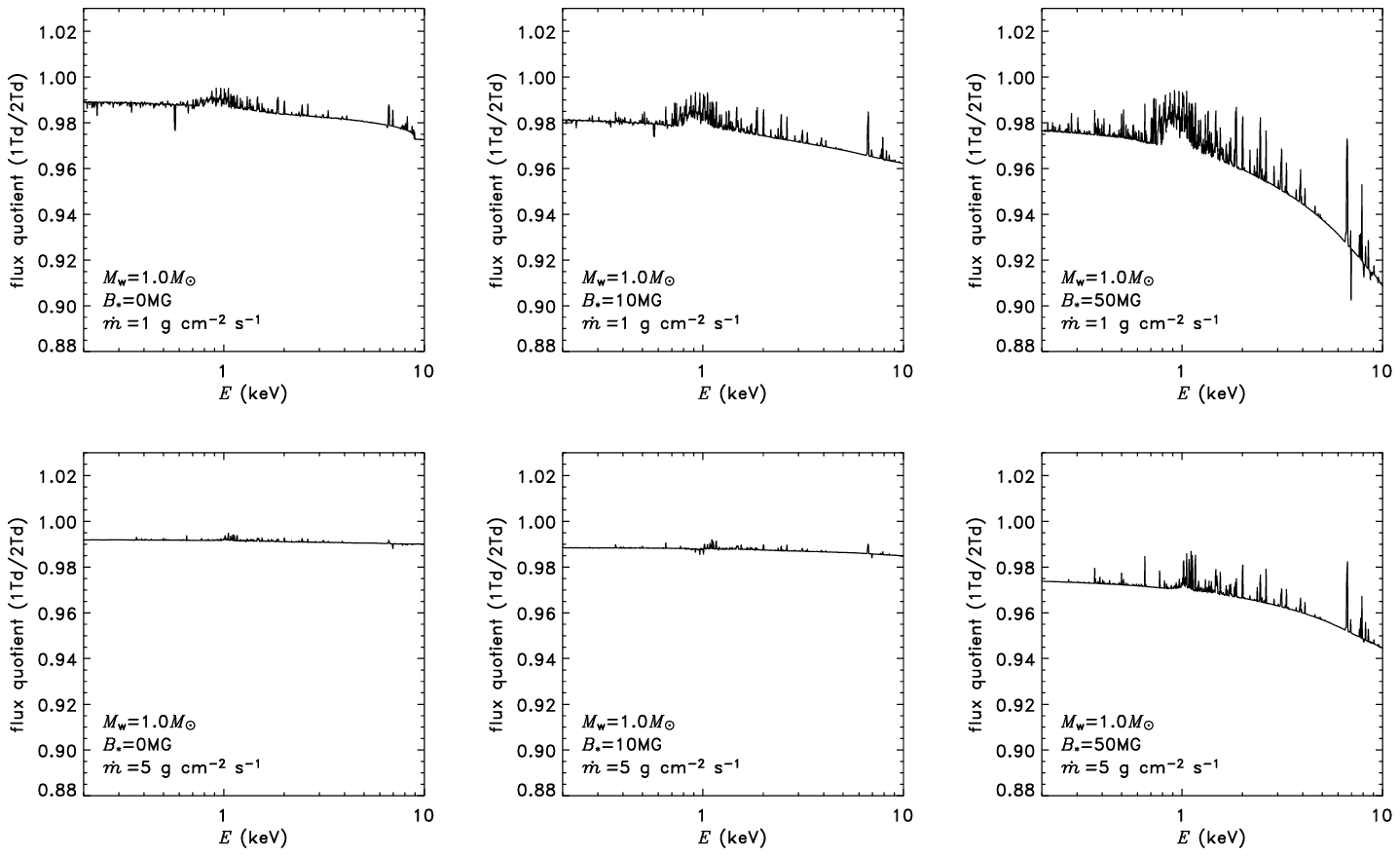}
\end{tabular}
\end{center}
\caption{
Quotients of spectra in 1T and 2T models (1Td/2Td),
as in Figure~\ref{fig.xquot.07} but with $M_{\rm w}=1.0~M_\odot$.
}
\label{fig.xquot.10}
\end{figure*}

The discrepancy between spectra predicted in the 1Td and 2Td models
  is more acute for systems with larger white-dwarf mass.
For cases with $M_{\rm w}=1.0\Msun$ 
  (see Figure~\ref{fig.xquot.10})
  the oversoftness of the 1Td predicted continuum
  is more pronounced in the $B_*=30$~MG case
  than in the comparable $B_*=50$~MG cases with $M_{\rm w}=0.7\Msun$.
The 1Td/2Td pattern of over- and under-prediction of line emission
  between $\sim 7$~keV and $\sim 9$~keV
  varies considerably with $B_*$ and $\dot{m}$
  (see the upper right panels of Figure~\ref{fig.xquot.10}).

\begin{figure*}
\begin{center}
\begin{tabular}{c}
\includegraphics[width=17.7cm]{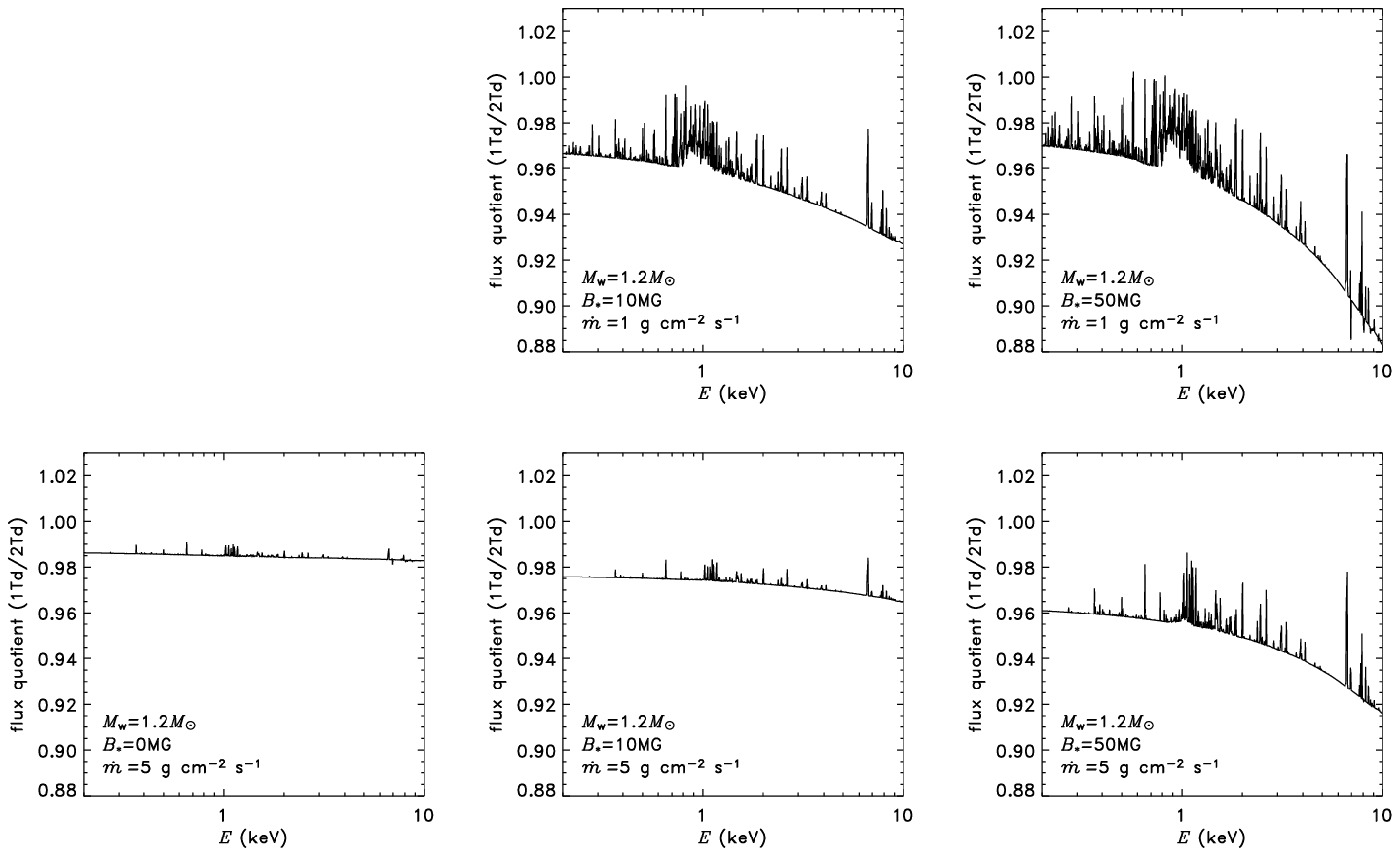}
\end{tabular}
\end{center}
\caption{
Quotients of spectra in 1T and 2T models (1Td/2Td),
as in Figure~\ref{fig.xquot.07} but with $M_{\rm w}=1.2~M_\odot$.
}
\label{fig.xquot.12}
\end{figure*}

For systems with a white dwarf with $M_{\rm w}=1.2\Msun$,
  the discrepancies are even greater than for $M_{\rm w}=1.0\Msun$ cases:
  the continuum around $\sim 10$~keV can be around 10\%
  underpredicted in the 1Td model,
  relative to the 2Td calculation
  (Figure~\ref{fig.xquot.12}).
A downward spike in the H-like Fe~K$\alpha$ emission
  that appears for $1.0\Msun$ cases
  (in Figure~\ref{fig.xquot.10},
  representing underprediction of this line by the 1Td model)
  is absent when $M_{\rm w}=1.2\Msun$,
  with all other parameters equal.

\subsubsection{Latitude dependence}

\begin{figure*}
\begin{center}
\begin{tabular}{c}
\includegraphics[width=17.7cm]{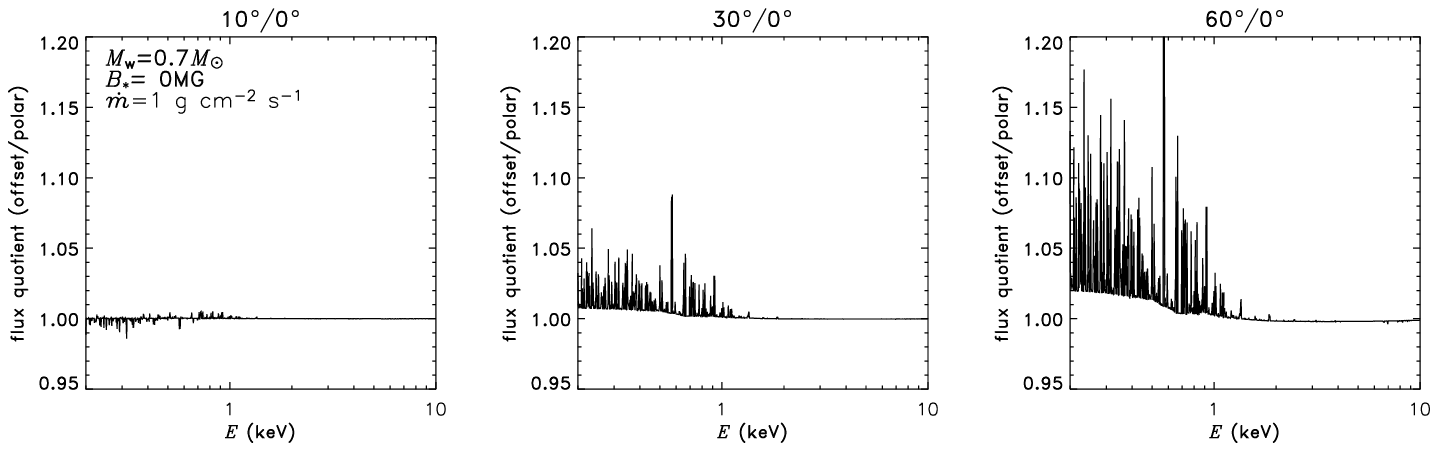}
\\
\includegraphics[width=17.7cm]{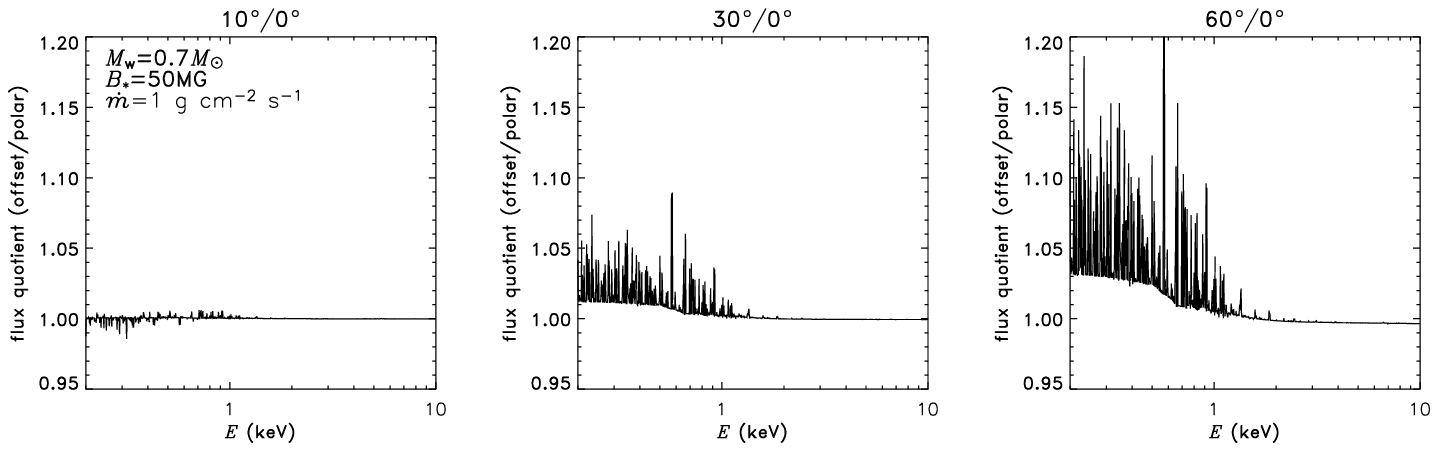}
\end{tabular}
\end{center}
\caption{
The effect of accretion latitude.
Quotients of spectra show different colatitudes of the accretion hot-spot:
   $10^\circ/0^\circ$, $30^\circ/0^\circ$ and $60^\circ/0^\circ$ 
   for left, middle and right columns respectively.
The upper panels show cases with weaker magnetic fields;
  the lower panels show cases with stronger fields.
In these cases the white-dwarf mass is $0.7~M_\odot$
and the specific accretion rate is $\dot{m}=1~\gram~\cm^{-2}~{\rm s}^{-1}$.
All else being equal, accretion further from the pole enhances
   the soft X-ray continuum and lines.
}
\label{fig.theta_M07m1}
\end{figure*}

The X-ray spectrum also varies with the location of the accretion spot
on the white-dwarf surface.
Latitudinal effects are slight when
the shock height is small compared to the stellar radius
(e.g. $M_{\rm w}=0.7\Msun$, $\dot{m}=5~\gram~\cm^{-2}~{\rm s}^{-1}$);
then we obtain $\la 2\%$ enhancements 
in the $\sim 1$~keV lines.
For the same white-dwarf mass
but lower accretion rate
(e.g.
$M_{\rm w}=0.7\Msun$, $\dot{m}=1~\gram~\cm^{-2}~{\rm s}^{-1}$),
a low-latitude accretion stream provides a softer X-ray continuum
than polar accretion.
Figure~\ref{fig.theta_M07m1}
shows the ratios of spectra:
for inclined accretion cases
($\theta_*=10^\circ, 30^\circ$ and $60^\circ$)
compared to vertical accretion onto the pole
($\theta_*=0^\circ$).
For low-latitude accretion,
lines are enhanced up to $\la 20\%$ at photon energies $\la 1$~keV.
In the cases with $M_{\rm w}=0.7\Msun$,
the size of $\theta_*$-dependency appears comparable for 
systems with low and high magnetic field strengths.

However in systems with a more massive white dwarf, $M_{\rm w}=1.2\Msun$,
the emission depends more sensitively on $\theta_*$,
for any field strength $B_*$.
In the cases with the shock at a relatively low radius
(e.g. $M_{\rm w}=1.2\Msun$, $\dot{m}=5~\gram~\cm^{-2}~{\rm s}^{-1}$)
the spectra are almost indistinguishable for $\theta_*=0^\circ, 10^\circ$;
the $\theta_*=30^\circ$ cases differ from $\theta_*=0^\circ$
by $\la 1\%$ in the $\sim 1$~keV lines.
At $\theta_*=60^\circ$ the enhancement of line emission is $\la 7\%$
when $B_*=10$~MG, but half as great when $B_*=50$~MG.
A similar variation with $B_*$ occurs
among cases with $M_{\rm w}=1.2\Msun$
and $\dot{m}=1~\gram~\cm^{-3}~{\rm s}^{-1}$
--- the greatest $\theta_*$-dependency
appears in the systems with the greatest relative shock height, $r_{\rm s}$.

\section{DISCUSSION}

\subsection{Summary for magnetic cataclysmic variables}

The incorporation of 2T effects in the post-shock accretion flows of mCVs 
  generally hardens the emitted keV X-ray continuum.
A dipolar accretion funnel further hardens the spectrum of a 2T flow,
  relative to models with simpler planar geometry.
A 2T dipolar model predicts a harder spectrum than a 2T planar model,
  if the white dwarf is as massive as $1\Msun$ or more.
The differences between planar and dipolar 2T accretion models 
  are greatest in systems where the shock height is large 
  compared to the stellar radius:
  i.e. greater $M_{\rm w}$ or lower $B_*$.
The spectral predictions of 1T and 2T dipolar models differ more significantly
  when the white dwarf is massive,
  has a strong magnetic field or low specific accretion rate.
Line emission is typically weaker around $\sim 1$~keV and $\sim 7-10$~keV
  in a 2T flow than in a 1T flow,
  except for the Fe~K$\alpha$ lines, which could become stronger.

This is qualitatively consistent with the 2T effects 
  found for basic planar accretion models
  without the inclusion of gravity in the hydrodynamic equations
  \citep{saxton2005};
Two-temperature effects upon X-ray spectra persist 
   despite the introduction of a dipolar magnetic field geometry,
   which in the 1T model of \cite{canalle2005} 
   soften the X-ray spectra relative to the results of \cite{cropper1999}.
This softening due to funnel geometry is, however, 
  smaller than the hardening apparent when 2T physics is incorporated.   
The 2Td model predicts the most hard photons for a given mass $M_{\rm w}$,
   and therefore its spectral fit to an observed spectrum
   will require a lower mass
   than the standard planar and 1T models do.

\subsection{Inferred masses}

\begin{figure*}
\begin{center}
\begin{tabular}{c}
\includegraphics[width=17.7cm]{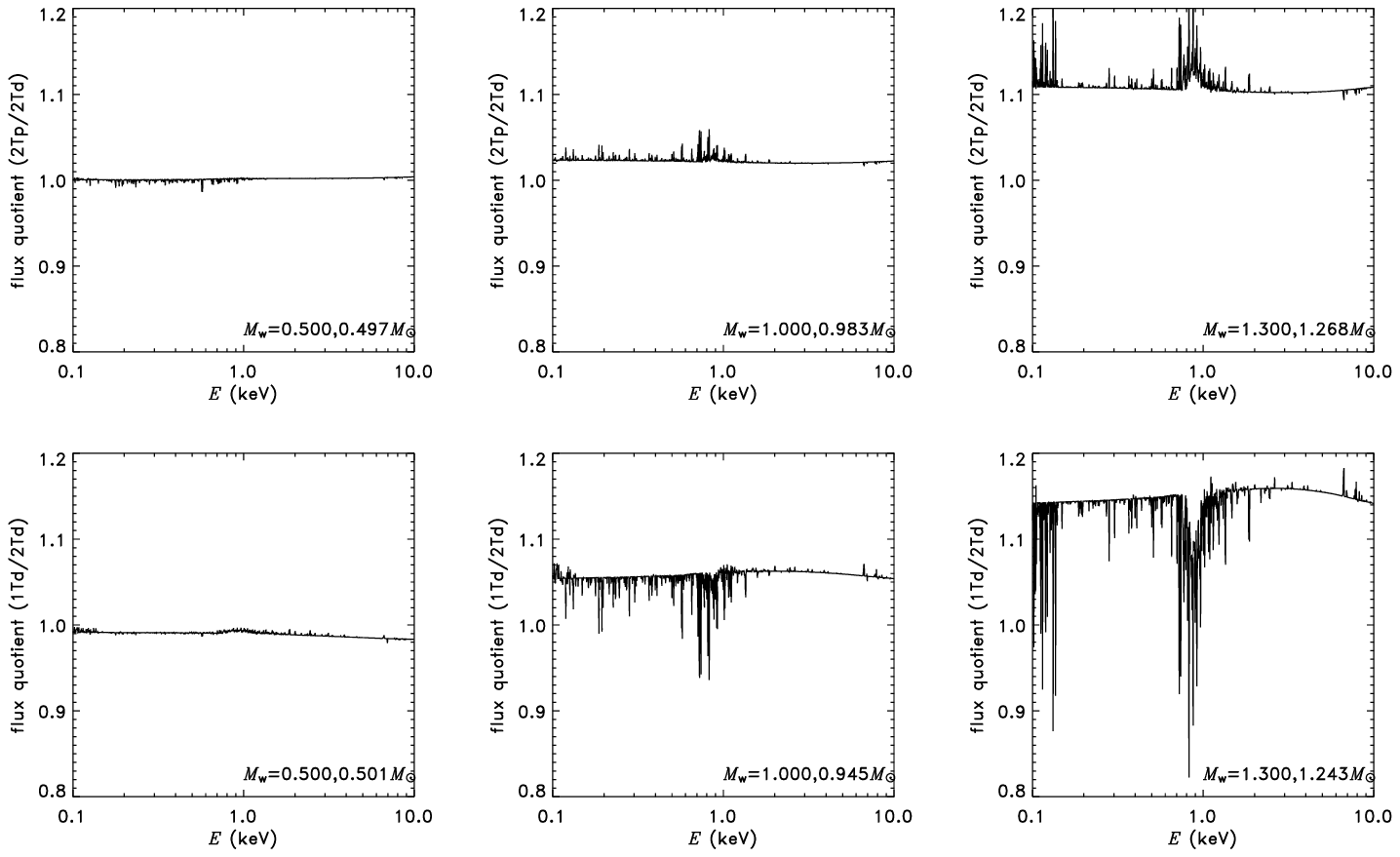}
\end{tabular}
\end{center}
\caption{
Matching the X-ray continua of two different accretion models
   requires different $M_{\rm w}$ values.
The top row compares 2T planar and dipolar models.
The bottom row compares 1T dipolar and 2T dipolar models.
Masses were adjusted to achieve approximately proportional continua
   (flat quotient profile).
We have fixed
   $B_*=50$MG,
   $\dot{m}=2~{\rm g}~{\rm cm}^{-2}~{\rm s}^{-1}$
   and $\theta_*=0^\circ$.
For low $M_{\rm w}$ (left panels) the inferred masses are similar;
   but for more massive cases
   the model masses differ appreciably.
}
\label{fig.massquotients}
\end{figure*}

To quantify how the mass estimates are affected,
   we calculated spectra for models with different $M_{\rm w}$,
   with the other parameters fixed
   ($\theta_*=0^\circ$,
   $\dot{m}=2~{\rm g}~{\rm cm}^{-2}~{\rm s}^{-1}$),
   for medium ($B_*=20$~MG)
   and strong-field ($B_*=50$~MG) cases.
For a given white dwarf mass,
   we manually match the shape of the X-ray continuum
   by adjusting the masses in 1Tp, 1Td, 2Tp and 2Td models
   \citep[][this paper respectively]{wu1994a,canalle2005,saxton2005}.
Each fit is defined by matching the ratio of spectra
   at 0.1~keV and 10~keV,
   with the flattest possible ratio at intermediate energies.
Figure~\ref{fig.massquotients} illustrates how
   the 2T and dipolar funnel effects
   change spectral inferences of the white dwarf mass ($M_{\rm w}$)
   from X-ray observations.
For low-mass white dwarfs ($M_{\rm w}\approx0.5M_\odot$)
   the 1T and 2T spectral predictions are similar.
However for cases with higher mass, say $M_{\rm w}\ga1.0M_\odot$,
   the spectra differ considerably (middle panels).

\begin{figure*}
\begin{center}
\begin{tabular}{cc}
\includegraphics[width=8.5cm]{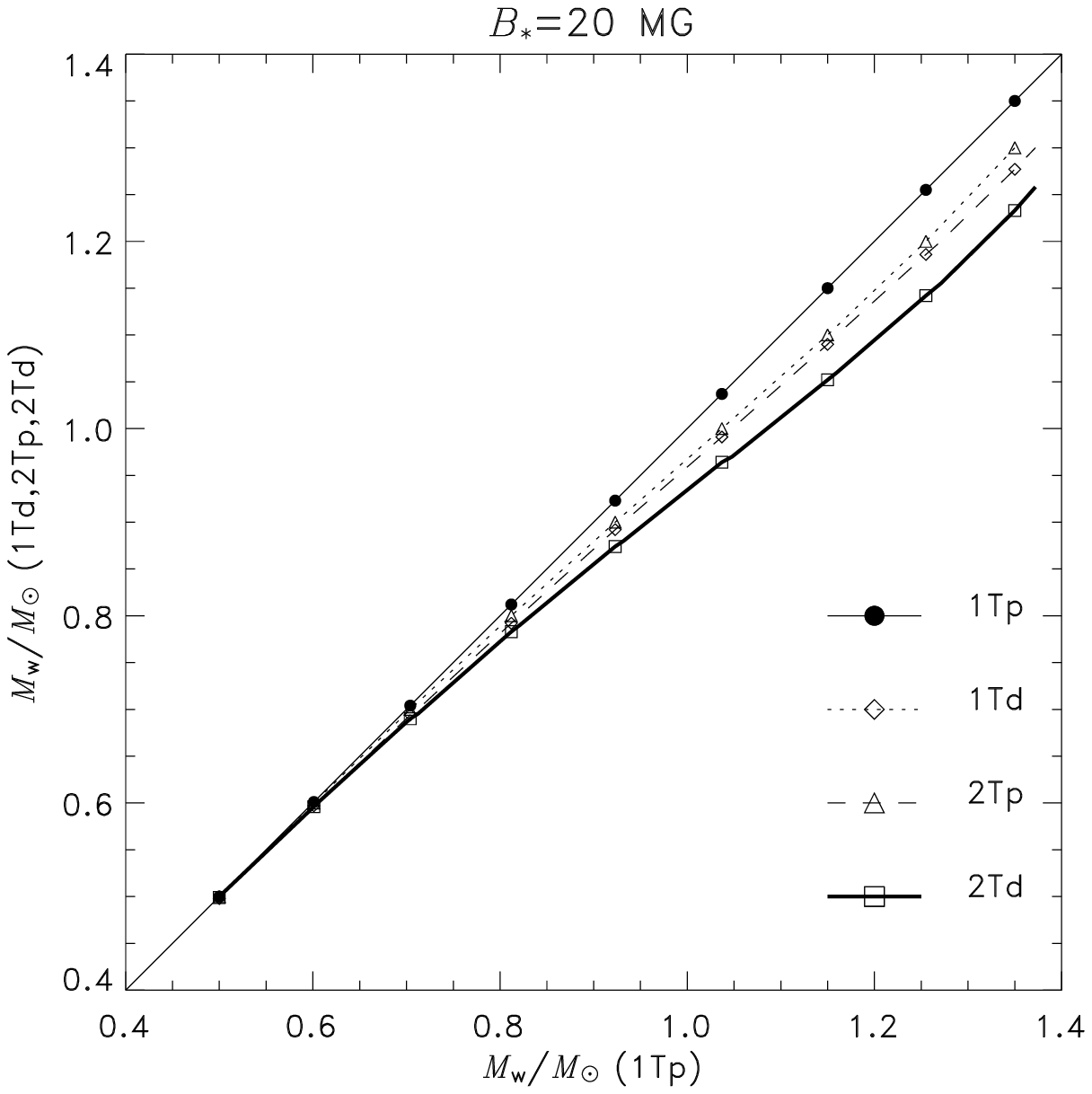}
&
\includegraphics[width=8.5cm]{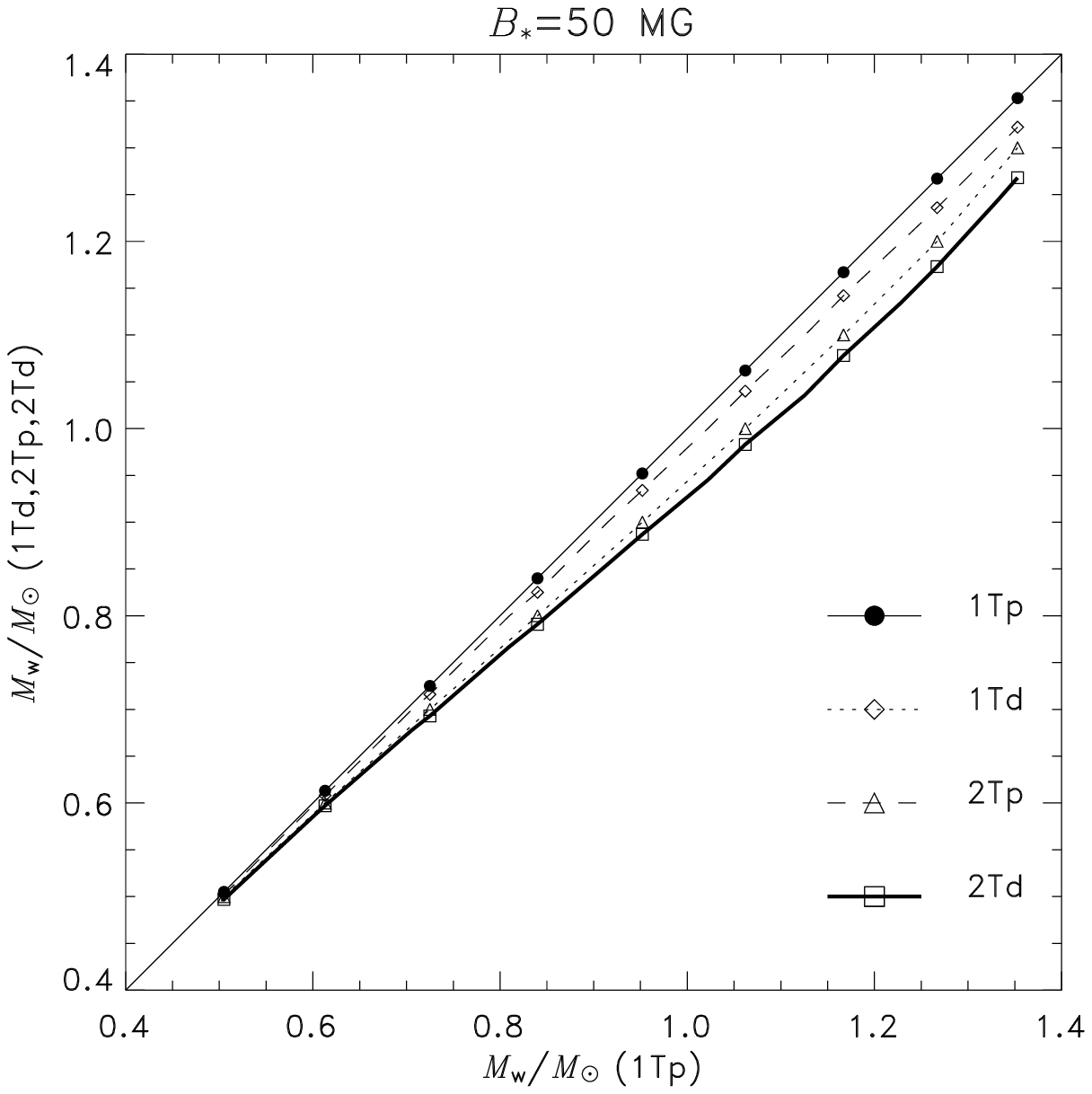}
\\
\end{tabular}
\end{center}
\caption{
Masses inferred by fitting continua of 1Tp, 1Td, 2Tp and 2Td models
to each other.
Radially variable gravity,
a dipolar accretion funnel
and two-temperature effects 
yield harder spectra for lower $M_{\rm w}$ values.
Here we have fixed the hot-spot size 
and $\dot{m}=2~\gram~\cm^{-2}~{\rm s}^{-1}$.
Symbols indictate $0.1\Msun$ steps in the 2Tp model;
the curves are
thin/$\bullet$ (1Tp),
dotted/$\triangle$ (2Tp),
dashed/$\Diamond$ (1Td)
and
thick/$\Box$ (2Td).
}
\label{fig.masses}
\end{figure*}

For large $M_{\rm w}$, the 2Tp models require lower masses
   than 1Tp models.
For $B_*=50$~MG
   the 1Td model requires greater masses than 2Tp models;
   but for $B_*=20$~MG the 1Tp model is more massive.
The 2T dipolar models consistently need the lowest $M_{\rm w}$
   (see corresponding fitted masses in Figure~\ref{fig.masses}).
The introduction of dipolar geometry to 2-temperature flows
   (from 2Tp to 2Td models) has hardened the spectra.
For example, a $1.00\Msun$ 2Tp model matches
   a 2Td model with $M_{\rm w}\approx0.98\Msun$
   (i.e. a 2\% discrepancy, see Table~\ref{table.mass}).
The X-ray continuum of a $1.00\Msun$ 1Td model
   is best matches a 2Td model with $M_w\approx0.95\Msun$
   (a 5\% discrepancy, see Figure~\ref{fig.massquotients}, lower middle panel).

We note that the flow geometry and 2T conditions
   can affect the line spectra as well as the continuum.
Here, we demonstrate that mass estimates obtained from the continuum fitting
   depends on the assumed geometry of the flow.
If the spectral line information were also considered,
   we would need to consider variations of the values of
   $B_*, \dot{m}, \theta_*$ and the hot-spot area too.
For the present study,
   it suffices to show that the joint inclusion of 2T phenomena
   and dipolar geometry
   can reduce mass estimates by $\la 9\%$
   near the Chandrasekhar limit,
   by $\la 7\%$ near a solar mass,
   but only $\la 1\%$ near $0.5~\Msun$.

We now consider the mass estimates of
   \cite{ramsay2000}
   who determined the mass of the white dwarf in 21 mCVs
   --
   8 polars and 13 intermediate polars (IPs)
   --
   using {\sl RXTE} PCA (2--60~keV) data.  
Using the 1Tp model prescribed by
   \cite{cropper1999},
   \cite{ramsay2000}
   found a mean mass of $0.87 \Msun$ and $0.80 \Msun$
   for his sample of IPs and polars respectively.
Applying a correction factor based on our results above
   gives a mean mass of $0.82 \Msun$ and $0.76 \Msun$
   for the IPs and polars respectively.

Other affects maybe important as well.
For instance,
   \cite{suleimanov2005}
   found that when they determined the mass of 14 IPs using
   {\sl RXTE} PCA plus HEXTE (10--200keV) data
   the mean mass of the IPs was reduced by $0.20 \Msun$ to $0.75 \Msun$.
This was explained by the fact that in IPs
   broad accretion curtains with high total absorption columns
   are thought to be present,
   which can affect the modelling of the continuum
   even at relatively hard X-ray energies.
Because polars have higher magnetic field strengths and
   hence more collimated accretion streams,
   absorption may affect their X-ray continua less than for IPs.
   
We note that the mean mass of white dwarfs in mCVs
   as determined by \cite{suleimanov2005}
   and in this paper is greater than that of
   isolated white dwarfs
   \citep[$0.56 \Msun$,][]{bergeron1992}.
We note that
   \cite{wickramasinghe2005}
   found that white dwarfs with high magnetic fields
   have a mean mass of $0.93 \Msun$
   -- significantly higher than that in isolated white dwarfs.
Our mass estimates remain consistent with the scenario
   in which magnetic white dwarfs are more massive than
   their non-magnetic counterparts.

\begin{table}
\centering
\caption{
Sets of $M_{\rm w}$ values that give comparable
spectral continua,
calculated for the 1Tp, 1Td, 2Tp and 2Td models.
The latter imply similar or lower masses.
For these mass fits, we fix
$\dot{m}=2$~g~cm$^{-2}$~s$^{-1}$
and $\theta_*=0^\circ$.
}
\label{table.mass}
\begin{center}
\begin{tabular}{cc}
\begin{tabular}{cccc}
\multicolumn{4}{c}{$B_*=20$~MG}\\
1Tp&1Td&2Tp&2Td\\
\hline
0.500&0.498&0.500&0.499\\
0.601&0.597&0.600&0.596\\
0.704&0.694&0.700&0.690\\
0.812&0.792&0.800&0.783\\
0.923&0.892&0.900&0.874\\
1.037&0.991&1.000&0.964\\
1.150&1.090&1.100&1.052\\
1.255&1.186&1.200&1.142\\
1.350&1.277&1.300&1.233\\
\\
\hline
\end{tabular}
\\
\\
\begin{tabular}{cccc}
\multicolumn{4}{c}{$B_*=50$~MG}\\
1Tp&1Td&2Tp&2Td\\
\hline
0.505&0.502&0.500&0.497\\
0.613&0.608&0.600&0.597\\
0.725&0.716&0.700&0.693\\
0.840&0.825&0.800&0.791\\
0.952&0.934&0.900&0.887\\
1.062&1.040&1.000&0.983\\
1.167&1.142&1.100&1.078\\
1.267&1.236&1.200&1.173\\
1.353&1.322&1.300&1.268\\
\\
\hline
\end{tabular}
\end{tabular}
\end{center}
\end{table}

\subsection{Further applications and scope}

Our 2-temperature formulation with magnetic funnel geometry
   is applicable to a variety of accreting stellar systems
   other than the cataclysmic variables.
With substitution of the relevant radiation processes,
   the model is applicable to accretion onto neutron stars,
and similarly it is applicable to field-aligned accretion onto T~Tauri stars.
The key assumptions are that the flow be hydrodynamic (collisional),
   field-aligned,
   and that radiation effects can be described in terms of local variables.

Our hydrodynamic formulation treats
   accretion flows that are steady and smooth:
   rapid temporal variability is omitted.
Continuous flows onto some mCVs oscillate at $\sim1$~Hz,
   interpreted as a thermal instability of the shock,
   \citep[see][]{langer1981,langer1982,chevalier1982,saxton1998}.
This is observed optically in only a minority of systems
   \citep[e.g.][]{middleditch1982,cropper1986,imamura1986,larsson1987,larsson1989,ramseyer1993,middleditch1997}
   but not yet in X-rays \citep[e.g.][]{wolff1999,christian2000}.
Thus fast oscillations may not affect X-ray spectra
   at presently observable levels.
The absence of a steady structure
   would preclude collisional ionisation equilibrium,
   which affects the spectral line strengths,
   as they critically depend on the density structure
   \citep[see e.g.][]{fujimoto1997,wu2001b}.
Our model is more applicable to cases with slower temporal variability --
   flaring which exceeds the post-shock free-fall time --
   by calculating sequences of steady models
   with different $\dot{m}$ values
   but identical $M_{\rm w}, B_*$ and $\theta_*$.

The spectral effects of spatial variations are harder to assess.
Mild inhomogeneities might be modelled by treating the accretion column
   as a distribution of streams with differing $\dot{m}$ and cross-sections,
   then summing their spectra.
By construction, we cannot model highly inhomogeneous flows:
   these would require highly resolved,
   time-dependent hydrodynamic simulations.
In some asynchronous accreting systems, a continuous flow
   may fragment into dense blobs due to a magnetic drag
   \citep[e.g.][]{arons1980,frank1988,king1993,wynn1995}.
However if the magnetic field is strong enough
   then the blobs shred into a continuous field-aligned flow
   as they near to the stellar surface
   \citep{arons1980}.
If the field is weak,
   blobs may survive to strike the stellar surface individually.
Such a shower is qualitatively different from 
   a pressure-supported stand-off shock.
However, emission from steady accretion shocks is universally observed in mCVs
   and is the major component,
   except at the softest energies
   where the emission from ballistic blobs contributes in most cases.
Spectral modelling of predominantly clumpy accretion
   is a challenge beyond the scope of this paper.


\section*{ACKNOWLEDGMENTS}

KW thanks TIARA for their hospitality during his visit there 
  and for support through a Visiting Fellowship. 
TIARA is operated under Academia Sinica 
  and the National Science Council Excellence Projects programs in Taiwan  
  administered through grant number NSC~94-2752-M-007-001. 

%
%
%


\def\jnl@style{\rm}
\def\aaref@jnl#1{{\jnl@style#1}}

\def\aaref@jnl#1{{\jnl@style#1\thinspace}}

\def\aj{\aaref@jnl{AJ}}                   
\def\araa{\aaref@jnl{ARA\&A}}             
\def\apj{\aaref@jnl{ApJ}}                 
\def\apjl{\aaref@jnl{ApJ}}                
\def\apjs{\aaref@jnl{ApJS}}               
\def\ao{\aaref@jnl{Appl.~Opt.}}           
\def\apss{\aaref@jnl{Ap\&SS}}             
\def\aap{\aaref@jnl{A\&A}}                
\def\aapr{\aaref@jnl{A\&A~Rev.}}          
\def\aaps{\aaref@jnl{A\&AS}}              
\def\azh{\aaref@jnl{AZh}}                 
\def\baas{\aaref@jnl{BAAS}}               
\def\jrasc{\aaref@jnl{JRASC}}             
\def\memras{\aaref@jnl{MmRAS}}            
\def\mnras{\aaref@jnl{MNRAS}}             
\def\pra{\aaref@jnl{Phys.~Rev.~A}}        
\def\prb{\aaref@jnl{Phys.~Rev.~B}}        
\def\prc{\aaref@jnl{Phys.~Rev.~C}}        
\def\prd{\aaref@jnl{Phys.~Rev.~D}}        
\def\pre{\aaref@jnl{Phys.~Rev.~E}}        
\def\prl{\aaref@jnl{Phys.~Rev.~Lett.}}    
\def\pasp{\aaref@jnl{PASP}}               
\def\pasj{\aaref@jnl{PASJ}}               
\def\qjras{\aaref@jnl{QJRAS}}             
\def\skytel{\aaref@jnl{S\&T}}             
\def\solphys{\aaref@jnl{Sol.~Phys.}}      
\def\sovast{\aaref@jnl{Soviet~Ast.}}      
\def\ssr{\aaref@jnl{Space~Sci.~Rev.}}     
\def\zap{\aaref@jnl{ZAp}}                 
\def\nat{\aaref@jnl{Nature}}              
\def\iaucirc{\aaref@jnl{IAU~Circ.}}       
\def\aplett{\aaref@jnl{Astrophys.~Lett.}} 
\def\apspr{\aaref@jnl{Astrophys.~Space~Phys.~Res.}}
\def\bain{\aaref@jnl{Bull.~Astron.~Inst.~Netherlands}} 
\def\fcp{\aaref@jnl{Fund.~Cosmic~Phys.}}  
\def\gca{\aaref@jnl{Geochim.~Cosmochim.~Acta}}   
\def\grl{\aaref@jnl{Geophys.~Res.~Lett.}} 
\def\jcp{\aaref@jnl{J.~Chem.~Phys.}}      
\def\jgr{\aaref@jnl{J.~Geophys.~Res.}}    
\def\jqsrt{\aaref@jnl{J.~Quant.~Spec.~Radiat.~Transf.}}
\def\memsai{\aaref@jnl{Mem.~Soc.~Astron.~Italiana}}
\def\nphysa{\aaref@jnl{Nucl.~Phys.~A}}   
\def\physrep{\aaref@jnl{Phys.~Rep.}}   
\def\physscr{\aaref@jnl{Phys.~Scr}}   
\def\planss{\aaref@jnl{Planet.~Space~Sci.}}   
\def\procspie{\aaref@jnl{Proc.~SPIE}}   

\let\astap=\aap
\let\apjlett=\apjl
\let\apjsupp=\apjs
\let\applopt=\ao



\appendix

\section{COORDINATE SYSTEM}
\label{app.coordinates}

Following
   \cite{canalle2005},
   we adopt a curvilinear coordinate system $(u,w,\varphi)$
   in which $\varphi$ is an azimuthal angle about the polar axis,
   $w$ is a path-length along a magnetic field line,
   and $u$ measures along an equipotential transverse to the field lines,
   effectively choosing the latitude of the field-line's footprint 
   on the stellar surface.
The dipolar coordinates correspond to conventional polar coordinates
   according to
\begin{equation}
	u={{\sin^2\theta}\over{r}}
	\ \ \mbox{and}
\end{equation}
\begin{equation}
	w={{\cos\theta}\over{r^2}}
	\ ,
\end{equation}
implying that
\begin{equation}
	w^2r^4+ur-1=0
	\ .
\end{equation}
We scale the coordinate axes so that
the stellar surface occurs at $r=1$.
The transformation to Cartesian coordinates obeys
\begin{equation}
	\left[{\begin{array}{cc}
		x\\y\\z
	\end{array}}\right]
	=
	\left[{\begin{array}{cc}
		\sqrt{u r^3} \cos\varphi 
		\\
		\sqrt{u r^3} \sin\varphi
		\\
		w r^3
	\end{array}}\right]
	\ .
\end{equation}
The orthogonal unit vectors,
  ($\BHu, \BHw, \BHphi$),
  vary spatially.
Locally they transform into their Cartesian counterparts
   ($\BHi$, $\BHj$, $\BHk$)
   according to
\begin{equation}
\left[\begin{array}{c}
  \hat{\mathbf u}\\
  \hat{\mathbf w}\\
  \BHphi
  \end{array}\right]
   = {\mathsf U} \left[\begin{array}{c}
  \BHi\\
  \BHj\\
  \BHk
   \end{array}\right]  \ .
\end{equation}
where (see e.g.\ \citealt{arfken})
  the transformation matrix is
\begin{equation}
  {\mathsf U}\!\equiv\!
\left[\begin{array}{ccc}
\frac{\displaystyle 1}{\displaystyle h_{1}}\frac{\displaystyle \partial x}{\displaystyle \partial u}     &
\frac{\displaystyle 1}{\displaystyle h_{1}}\frac{\displaystyle \partial y}{\displaystyle \partial u}     &
\frac{\displaystyle 1}{\displaystyle h_{1}}\frac{\displaystyle \partial z}{\displaystyle \partial u}\\
\frac{\displaystyle 1}{\displaystyle h_{2}}\frac{\displaystyle \partial x}{\displaystyle \partial w}     &
\frac{\displaystyle 1}{\displaystyle h_{2}}\frac{\displaystyle \partial y}{\displaystyle \partial w}     &
\frac{\displaystyle 1}{\displaystyle h_{2}}\frac{\displaystyle \partial z}{\displaystyle \partial w}\\
\frac{\displaystyle 1}{\displaystyle h_{3}}\frac{\displaystyle \partial x}{\displaystyle \partial \varphi} &
\frac{\displaystyle 1}{\displaystyle h_{3}}\frac{\displaystyle \partial y}{\displaystyle \partial \varphi} &
\frac{\displaystyle 1}{\displaystyle h_{3}}\frac{\displaystyle \partial z}{\displaystyle \partial \varphi}
  \end{array}\right] \ .
\end{equation}
The coefficients $h_{1}$, $h_{2}$ and $h_{3}$
  comprising the metric of the curvilinear coordinates system,
  are defined by
\begin{equation}
  h_{[1,2,3]}^2 \equiv 
	 \left(\frac{\partial x}{\partial [u,w,\varphi]}\right)^{2}
         +\left(\frac{\partial y}{\partial [u,w,\varphi]} \right)^{2}
         +\left(\frac{\partial z}{\partial [u,w,\varphi]}\right)^{2}   \ ,
\label{h1 definition}
\label{h2 definition}
\label{h3 definition}
\end{equation}
with explicit functional forms given in \cite{canalle2005}.
For a small fluid element following a magnetic field line,
its transverse area varies as $h_1h_3$
in response to the converge of the field.
The rate of convergence of the field lines,
and thus the lateral compression of field-channelled inflow,
can be expressed in terms of a function
\begin{equation}
	{\mathcal H}\equiv
	{\partial\over{\partial w}} \ln(h_1 h_3)
	\ .
\end{equation}

\section{NUMERICAL SOLUTION}
\label{app.solver}

We find that it is convenient to express the post-shock structure
in terms of a specific entropy variable,
\begin{equation}
s\equiv P\rho^{-\gamma}
=\left({
	{{h_1h_3}\over{C}}
}\right)^{\gamma-1}
(\xi-v) v^\gamma
\ ,
\label{eq.def.entropy}
\end{equation}
which has a gradient equation,
\begin{equation}
	{{ds}\over{dw}}
	=
	-{{(\gamma-1)\tilde\Lambda s}\over{(\xi-v)v^{5/2}}}
	=
	-(\gamma-1)\left({{h_1h_3}\over{C}}\right)^{\gamma-1}
	\tilde\Lambda v^{\gamma-5/2}
	\ .
\label{eq.dsdw}
\end{equation}
Since the expression (\ref{eq.dsdw}) never changes sign,
the spatial variation of $s$ is guaranteed to be monotonic
throughout the post-shock flow.
The maximum value of $s$ occurs at the shock;
the specific entropy falls to zero at the stellar surface.
Thus $s$ is a convenient replacement for $w$ 
in the role of independent, integration variable.
Differential equations in terms of $ds$ are obtained
by multiplying $dw/ds$ into
(\ref{eq.dxidw}), (\ref{eq.dvdw}), (\ref{eq.dPedw}) and (\ref{eq.dsigmadw})
and then simplifying algebraic terms
to reduce numerical round-off errors.

The strong shock boundary conditions imply that
\begin{equation}
 v_{\rm s}={1\over{4\sqrt{ R_{\rm s} }}}
\end{equation}
and
\begin{equation}
 \xi_{\rm s}={1\over{\sqrt{R_{\rm s}}}}
 \ ,
\end{equation}
\citep{canalle2005},
where we use a unit convention based on
$C=1$,
the stellar radius $R_{\rm w}=1$
and superficial escape velocity $V_{\rm w}\equiv\sqrt{2GM_{\rm w}/R_{\rm w}}=1$.
The shock values of $s$ and $P_{\rm e}$ follow from 
(\ref{eq.def.entropy}) and (\ref{eq.def.Pe}),
with $\sigma_{\rm s}$ treated as a free parameter.
Assuming trial estimates of the shock location,
we integrate inwards to $s\rightarrow 0$
and check for consistency between this inner radius
and the stellar surface, $r=1$.
A numerical root-finder iterates to a value of $R_{\rm s}$
that matches both boundary conditions.




\bsp

\label{lastpage}

\end{document}